\documentclass [prb aps] {revtex4-1}
\usepackage{graphicx}
\usepackage{amsmath}
\usepackage{amssymb}
\usepackage{color}

\newcommand{\p}{\psi }

\newcommand{\be}{\begin{equation}}
\newcommand{\ee}{\end{equation} }
\newcommand{\ba}{\begin{eqnarray} }
\newcommand{\ea}{\end{eqnarray} }
\newcommand{\n}{\nonumber \\ }

\begin{document}

\title{The devil's staircase in 1-dimensional dipolar Bose gases in optical lattices}

\author{F. J. Burnell}
\affiliation{Department of Physics, Princeton University, Princeton,
New Jersey 08544, USA} %

\date{\today}

\begin{abstract}
We consider a single-component gas of dipolar bosons
confined in a one-dimensional optical lattice, where the dipoles are
aligned such that the long-ranged dipolar interactions are maximally
repulsive. In the limit of zero inter-site hopping and sufficiently
large on-site interaction, the phase diagram is a complete devil's
staircase for filling fractions between 0 and 1: every
commensurate state at a rational filling is stable over a finite
interval in chemical potential, and for every chemical potential the system 
is in a gapped commensurate phase. We perturb away from this limit in
two experimentally motivated directions involving the addition of
hopping and a reduction of the onsite interaction. The addition of
hopping alone yields a phase diagram, which we compute in
perturbation theory in the hopping, where the commensurate Mott
phases now compete with the superfluid.  
We capture the physics of the Mott-superfluid phase transitions via bosonization.
In a finite trap, we argue using an LDA
and simulated annealing that this results in 
regions of commensurate states separated by patches with dominant superfluid correlations. 
Further softening of the
onsite interaction yields alternative commensurate states with
double occupancies which can form a devil's staircase of their own; we describe these states
and discuss the possible transitions between them.  Adding a hopping term in this case 
produces one-dimensional ``supersolids'' which simultaneously exhibit
discrete broken symmetries and superfluidity.  The contents of this work was first 
published as a chapter in the doctoral thesis `On Exotic Orders in Strongly Correlated Systems' (Princeton University, 09/09, supervised by S. L. Sondhi), and constitutes a considerably more detailed version of [PRB 80: 174519 (2009)].
\end{abstract}

\maketitle

\section{Introduction}

When commensuration effects compete 
with long-ranged interactions, a startling richness of phases can arise.  This 
interplay between long-ranged forces and lattice effects has been studied
in a variety of models -- from the Ising model \cite{Hubbard} to interfaces between crystal surfaces \cite{Schulz}.  
In this work, we explore a new context in which such physics leads to 
rich mathematical structure -- that of the cold dipolar Bose gas.  
Recent
advances in laser cooling and trapping have opened the possibility of 
creating cold Bose gases in optical lattice potentials.  
If these bosons have sufficiently strong dipolar interactions, the phase 
portrait is controlled by the interplay between the infinite-ranged 
dipolar repulsion\footnote{The dipolar interaction can be made repulsive everywhere on the line 
by polarizing the dipoles with an external field}, and commensuration effects due to the optical lattice --
giving a potential experimental realization of the diverse phases 
expected in such a system.  Furthermore, considering which parameters can be 
easily tuned in such experiments naturally opens questions about the ground 
states of these systems in new regions of parameter space. 

The driving force behind the strikingly rich phase diagram in these 
systems is that long ranged interactions on a lattice can stabilize 
exceptionally intricate ground-state structures in the
classical, or strong coupling, limit. A particularly elegant example
of this is the case of classical particles in a one-dimensional (1D)
lattice interacting via an infinite-ranged convex potential studied
by Pokrovsky and Uimin, and Hubbard
(PUH)~\cite{Hubbard,PokrovskyUimin,Pokrovsky}. Here, it can be shown
that the ground state filling fraction as a function of chemical
potential $\mu$ is a complete devil's staircase~\cite{Bak}, in which
every rational filling fraction between 0 and 1 is stable over a
finite interval in $\mu$, and the total measure of all such
intervals exhausts the full range of $\mu$.

This devil's staircase of PUH has dramatic
consequences for the physics of quasi 1D cold atomic gases. Building
on the existing understanding of this classical limit, we consider
two perturbations of the devil's staircase that arise naturally in
the experimental setting of cold atomic gases. The first of these is
the introduction of a quantum kinetic energy, due 
to the finite depth of the optical lattice, which now renders the
problem sensitive to particle statistics.  We focus on the
bosonic case, as this case is most likely to be realized 
experimentally.  Tuning away from the classical limit
has the well understood effect of initiating a competition between
the crystalline, Mott phase that exists at zero hopping and the
superfluid (Luttinger liquid in $d=1$) that must exist at all
fillings at sufficiently large hopping.  The phase diagram we 
find is thus an extremely complex variant of the usual Hubbard model 
story, with an infinite number of Mott-Hubbard lobes corresponding to
commensurate phases.  

The second perturbation involves tuning the
onsite interaction independently of the dipolar potential.
This allows for a controlled departure from convexity, and generates 
new classical states comparable in complexity to the 
PUH states considered previously.  
This introduces doubly occupied
sites in the classical limit. While describing the resulting phase
diagram in complete and rigorous detail is beyond the scope of this 
project, we give an account of the ``staircase''
structure of the initial instability and of regions of the phase
diagram where the classical states exhibit superlattices of
added charge built on underlying PUH states. At least some of these
regions exhibit devil's staircases of their own. Finally, upon
introducing hopping we are led to an infinite set of
``supersolids''---which in this context are phases that are both
Luttinger liquids and break discrete translational symmetries.

The structure of this paper is as follows.  We begin in Section \ref{ModelSect} with an overview of the model and a sketch of the relevant experimental parameters.  
\ref{ClassSect} gives a detailed description of the intriguing infinite family of classical
solutions to this model, described in \cite{Hubbard, Pokrovsky}, and discuss the fractal
structure known as the devil's staircase which constitutes the classical
phase portrait.  Then in Section \ref{PertSect},
 we use a strong coupling perturbation theory similar to that previously studied in the
Bose-Hubbard model~\cite{FreericksMonien}, and in extended
Bose-Hubbard models with nearest-neighbor
interactions~\cite{WhiteMonien, ScalletarBH1},
to calculate the boundaries of the Mott-Hubbard lobes where these commensurate
classical states liquify.  We also review how bosonization techniques admit a 
Luttinger liquid description of these phase transitions.  This approach proves
helpful in understanding the expected density profiles of atoms in a parabolic
trapping potential -- a question we address for both quantum and classical
models in the convex limit in Section \ref{FiniteTrapSect}.

In Section \ref{ConvexSect}, we switch gears and evaluate the classical
phase diagram as $U_0$ is tuned away from convexity.  This reveals a mind-bogglingly
complex array of possible phases, only some of which we are able to describe
here.  However, one class of such phases, in which the ground states resemble a two-component
version of the HUP ground states, are of particular interest.  Section \ref{SSSect}
discusses in more detail the classical structure of these phases, as well as
their behavior at finite hopping -- where we find super-solid like phases.
We have relegated several detailed calculations to Section \ref{Appendix}, which
contains supplementary material.

\section{ The ultra-cold dipolar Bose Gas} \label{ModelSect}

The unprecedented control over experimental parameters in trapped
ultracold atomic gases has substantially widened the range of
possible exotic phases of matter that can be explored. Optical
lattices can be used to simulate simple lattice models and/or vary
the system dimensionality, while the interatomic interactions can
be varied via a magnetically-tunable Feshbach
resonance~\cite{bloch2008}. Thus far, the focus in these systems has largely been on
\textit{contact} interactions, since these generally provide a
good description of atom-atom scattering in the low energy limit.
However, with the realization of atomic gases with strong magnetic
dipole moments~\cite{lahaye2007} and the prospect of working with
molecules that exhibit electric dipole
moments~\cite{sage2005,Ni2008}, there is now substantial interest in
examining the physics of long-ranged interactions in these
systems\footnote{For a review of the current understanding of dipolar interactions, and the parameter r{\'e}gime currently accessible to experiments, see ~\cite{DipoleReview, DipoleReview2}}.  In this section we will present the model, motivated by 
these new experimental possibilities, which we will discuss in the 
remaining sections of this paper.

\subsection{Bosons in 1D optical lattices}

We begin by discussing the basic physics of dipolar molecules in an
optical lattice.  Experimentally, the scenario is as follows:
atoms, or in some cases even molecules,
 can be trapped using spatially varying electric fields.  In optical traps, 
this frequency can be tuned to be close to an atomic transition.  In this case the 
atoms are highly polarizable at this frequency, and hence feel a relatively 
strong trapping potential.  A lattice potential can then be made  for the gas of atoms, by
using counter-propagating laser beams
to create a periodic pattern of maxima and minima of the electric field's
intensity.
The depth of the lattice is tuned by the intensity of the laser light;
the lattice spacing is determined by the wavelength of the standing
light-waves.

To explore 1D physics, in practice a 3D optical lattice is generated, but
with a lattice depth and spacing much greater in the first two dimensions
than in the third.  This effectively creates an array of 1D tubes, with a
relatively weaker periodic potential and shorter inter-particle spacing 
along each tube.  Typically, the whole ensemble
of tubes is also confined in a long parabolic trap, to ensure that the particles
remain inside the optical lattice system.

Here we will ignore the complications due to interactions between tubes.  To
simplify the discussion, we will also begin by omitting the parabolic
trapping potential -- we will return to its effects in Section \ref{FiniteTrapSect}.
Hence the Hamiltonian we study is:
 \ba \label{HHup}
 H &= &V_0 \sum_{i< j}\frac{1}{r_{ij}^3} n_i n_j + \frac{U}{2} \sum_{i} n_i (n_i-1) \n
 &&- \mu \sum_i n_i -t \sum_{i} c^\dag_i c_{i+1} +h.c.
 \ea
This model describes bosons in a deep 1D optical lattice with
hopping amplitude $t$, on-site interaction energy $U$, and an
infinite-ranged dipole-dipole interaction $V_0/r^3$.
Since the hopping potential $t$ is controlled by the depth of the
optical lattice, it can be tuned over a wide range of values.
In the cases where the bosons are atoms, e.g.\
$^{52}$Cr~\cite{lahaye2007}, the on-site interaction $U$ may also be
easily tuned using Feshbach resonances. We set the dipolar
interaction to be maximally repulsive by aligning the dipoles
perpendicular to the 1D chain.  We also note that though dipolar
interactions will couple the different tubes, aligning the dipoles at an angle 
such that $\cos (\theta) = 1/\sqrt{3}$ cancels this interaction in
one of the two remaining dimensions.  Hence to access the 1D r{\'e}gime,
one would ideally work with a single 2D array of optical tubes.

Though our focus here is on the realm of possibly, rather than currently,
attainable states of the dipolar Bose gas, it is instructive to ask
what range of parameters can be attained in current experiments. The
critical ratios are $t/V_0$ and $U_0/V_0$.

In cold atomic gases, there are essentially two ways to generate 
dipolar interactions: by cooling atoms, such as $^{52}$ Cr, with large 
magnetic dipole moments, or by creating molecules with electric dipole 
moments -- typically by generating heteronuclear molecules which 
have an intrinsic dipole moment, for example, with $^{41}$K $^{87}$Rb.  Techniques for 
cooling and trapping atoms with large magnetic dipole moments have already been 
developed \cite{Lahaye07};  cooling dipolar molecules is significantly more 
difficult, though significant progress towards creating cold molecules in their 
2-body ground states has been achieved, most 
notably in fermionic systems \cite{Jin, Ospelkaus}.  Since
the experimentally attainable electric
dipole moments are much larger than their magnetic counterparts, 
it is interesting to consider the parameter r{'e}gimes accessible in  both
types of experiments.

Current ultracold bosonic molecules have electric dipole moments
of the order of $d \approx 1$ Debye; for $^{52}Cr$ and $^{87}Rb$ the 
magnetic dipole moments are 
$6 \mu_B$ and $1 \mu_B$, respectively.  Using these values, 
the dipolar interaction
strength at a distance of $n$ lattice spacings is 
\ba
 V_{dip}^{El} &= &
\frac{1}{n^3} \frac{(d D)^2}{4 \pi \epsilon_0 a^3} \ \ =  \frac{d^2
(3.338 \times 10^{-30} C m )^2}{4 \pi 8.854 \times 10^{-12} C^2/
(N m^2) a^3} \frac{d^2}{n^3} \n &\approx & \frac{1}{n^3 a^3}  1.0
\times 10^{-22} N m \n
V_{dip}^{Mag} &=&
\frac{1}{n^3} \frac{\mu_0 (\mu \mu_B)^2}{4 \pi a^3} \ \ =  \frac{
(4 \pi \times 10^{-7} N A^{-2} \mu^2 (9.27 \times 10^{-24} A m^2 )^2}{4 \pi a^3} \frac{1}{n^3} \n 
&\approx & \frac{\mu^2}{n^3 a^3}  8.6
\times 10^{-27} N m 
\ea 
where $a$ is the optical lattice spacing
in $nm$, $d =0.6$ is the electric dipole moment measured in Debye, and $\mu = 6.0$ for $^{52}Cr$ and $1.0$ for $^{87}Rb$ is the magnetic dipole 
moment in Bohr magnetons.  

To find the hopping parameter requires calculating the overlap integrals
of Wannier functions as a function of the depth of the optical lattice.
The result is that the hopping $t$ is given by \cite{DupuisRev}:
\be
t = \frac{4 E_r}{\sqrt{\pi}} \left (\frac{A}{E_r} \right) ^{3/4} e^{-2 \sqrt{ A /E_r}}
\ee
where  $A $ is the intensity of the laser beam,
and $E_r$ is the recoil energy of a single atom, $E_r = h^2 / (8 m a^2)$.  Current
experiments can attain lattice spacings of order $a = 500 nm$, at laser intensities on
the order of $10- 20$ recoils.  This gives a range of hopping parameters listed in Table
\ref{HoppingTable}.  We see that, assuming that the electric dipole moments of the 
molecules do not significantly alter the facility with which they can be trapped (a 
reasonable assumption since the frequency of the molecular transitions is significantly lower than that of the atomic transitions to which the laser light is tuned\footnote{Recall that 
the optical trap is created using laser light that is tuned to be close to an atomic 
transition, so that the atoms are highly polarizable at this frequency.}), fractions $t /V_0 ~ 10^{-2}$ are well within the range of experiments on polar molecules.  For magnetic dipole moments, the numerical value of $t/V_0$ is larger by a factor of $300$ (for $^{52}$Cr) to $6000$
(for $^{87}$Rb), and hence we do not expect to see any Mott physics associated with these states 
in currently realistic trapping potentials.

\begin{table}[h]  \label{HoppingTable}
\begin{center}
\begin{tabular}{|c c|c|c|c|}
\hline
& $a (nm)$ & $300$ & $400$ & $500$  \\
\hline
$A/E_r$ & &   & & \\
\hline
5  & & $0.184$ & $0.25$  & $0.306$ \\
10 & & $0.048$  & $0.065$  & $0.081$ \\
15 & & $0.016$  & $0.021$  & $ 0.026$ \\
20 & & $0.006$  & $0.008$ &  $ 0.010$ \\
\hline
\end{tabular}
\end{center}
\caption[Experimental values of dipolar interaction strength for polar molecules]{ Experimentally attainable values of the hopping $t/V_0$
as a function of the lattice constant and laser intensity in the
approximate range of experimentally realizable values.  Here we
use parameters relevant to polar molecules -- an electric dipole moment of $0.6$ Debye, and the mass of $^{41}$K $^{87}$Rb.  The corresponding values for $^{52}$Cr are larger by a factor of $\approx 300$. }
\end{table}

The effect of deforming the on-site repulsion $U$ away from
convexity is rather more difficult to probe experimentally. To
attain this r{\'e}gime, Feshbach resonances must be used to tune
the scattering length to be relatively large and negative, to
compensate for the strong dipole-dipole interaction felt by two
bosons confined to the same lattice site.  Calculating the overlap
of the Wannier wave functions in a deep optical lattice, 
where the particles are effectively localized at a single 
lattice site, gives
 \be
U = \frac{\sqrt{ 8\pi} a_s}{a} E_r \left( \frac{A}{E_r}
\right)^{3/4} \ \ \approx 5 \times 10^{-29} \frac{a_s}{a} N m
 \ee 
where $a_s$ is the
scattering length of the bosons, tuned by Feshbach resonances, and
we take $\frac{A}{E_r} = 20, a = 500 nm$.  To this we must add the effective dipolar
interaction, whose approximate order of magnitude we obtain by
noting that two bosons confined to the same site are a separation
of no more than $a/2$ apart -- giving a dipolar energy of
order $2 \times 10^{-30} N m$.  Hence a priori one would expect that on-site
interactions are easily tuned away from convexity by making the
scattering length $a_s$ sufficiently negative.  
In an optical lattice a reasonable limit \cite{bloch2008} is 
$a_S < a$, so some tuning away from convexity may be possible\footnote{
We are not aware of existing calculations which account for both 
scattering and dipolar interactions in an optical lattice; hence 
this estimate is obtained by assuming that the relevant parameter is 
the effective scattering length due to both scattering and dipolar interactions,
which we take to be positive.}.

\section{Classical solutions and the devil's staircase} \label{ClassSect}

The notion that longer-ranged interactions can stabilize
fractional fillings in the Mott state is well-known in the study
of extended Hubbard models. With only on-site interactions, 
the Hubbard model itself has commensurate states only at integer fillings; extended Hubbard models with repulsion between neighboring sites
can stabilize such states at half filling, and so on. 
The HUP case is simply the
extreme limit of the extended Hubbard model: convex interactions
favor arrangements in which particles are spread out as
homogeneously as possible given the filling fraction.  In the case
of infinite-ranged convex interactions, this yields a constraint
that must be satisfied at all distance scales, leading to a
pattern that is unique, up to global lattice transformations, for
every filling fraction.  Hence the classical configurations are
devoid of {\em all} local degeneracies, and commensurate structure
exists at all length scales.

In our model Hamiltonian (\ref{HHup}), if $U$ is sufficiently large
 the potential is everywhere
convex -- meaning that
$$
 V(x) \leq
\lambda V(x-1+\lambda) + (1-\lambda) V( x+ \lambda )
\mbox {   for  }0 \leq \lambda \leq 1 \ \ .
$$
The $t=0$ (classical) ground states of (\ref{HHup}) in this case are therefore
those of
a PUH Hamiltonian: for every rational filling fraction $\nu= p/q$,
the ground state is periodic with period $q$~\cite{Hubbard}.  Each
such ground state is unique up to global lattice
translations~\cite{Burkov}. We denote these states {\it commensurate
ground states} (CGS). Adding or removing a single particle from a
CGS in the infinite volume limit produces a {\it
 q-soliton state} (qSS) containing $q$ fractionally charged solitons
of charge $1/q$.  We now review the nature and energetics of
the CGS and qSS states.

\subsection{ Commensurate Ground States}

We now review the detailed construction of the CGS classical
ground states. To do so, it is convenient to characterize a state
by the set of all of its inter-particle distances. Call particles
$l^{th}$ neighbors if there are $l-1$ {\em occupied} sites
lying between them (their separation will generally exceed this).  
Any static configuration of particles on
the lattice is characterized by the set ($\mathcal{S}$) of sets
$S_l \equiv \{ r_l^{(1)}, r_l^{(2)}, ...r_l^{(N)} \} $ of $l^{th}$ neighbor distances
between a particle and its $l^{th}$ neighbor to the right.  (On a finite-sized chain it is 
simplest to consider periodic boundary conditions, in which case $r_l^{(i)}$ exists for 
all $i$ and $l< N$.)  
The energy of such a configuration is
 \be 
E = \sum_l \sum_{i} V(r_l^{(i)})
 \ee
 where $V(r)$ is the (convex) potential.  Here the
second sum is a sum over individual particles; the first sum is a
sum over all neighbors of each particle. A solution which
minimizes the inner sum for each $l$ also minimizes the energy;
hence it suffices to find the set $S_l$ for which
 \be
\label{Eq.En} E_l = \sum_i V(r_l^{(i)}) \ee 
 is minimal, and
establish that a configuration for which $S_l$ has this form for
all $l$ does indeed exist.

\subsubsection{ Energetics of convex interactions}

The first step in constructing the HUP ground states is to
establish which $S_l$, for a given $l$, minimizes the energy
(\ref{Eq.En}).   For convex interactions, the optimal $S_l$ is a
{\it maximally compact} set -- e.g. the one for which all $r_l$
are as close as possible to the same value.  This is commensurate
with the intuition that convex interactions favor maximally
uniform distributions of particles, and we will prove presently
why this is so.

For a given $l$ and filling fraction $\nu$, what is the maximally compact $S_l$?  In the
absence of a lattice, maximal compactness would imply $r_l^{(i)} \equiv r_l$ for all
$i$; however on a lattice we must also allow $S_l = \{ r_l, r_l +1
\}$.  To see why, recall that the sum of all distances between first neighbors
must be the total length of the lattice -- or, more generally, that for a lattice of length $L$ we have
\be
\sum_i r_l^{(i)} = l L  .
\ee
The average separation between $l^{th}$ neighbors is thus:
\be
\bar{r _l} = \frac{1}{N}\sum_i r_l^{(i)} = l /\nu
\ee
where $\nu = N/L$.  For rational
fillings $\nu = p / q $, this gives:
\be
\bar{r_l} = l q /p
\ee
Thus either $p | l$ and the maximally compact set is $S_l =
\{\bar{r_l}  \} = \frac{l}{p} q$, or $\bar{r_l}$ is not an integer, and hence
$S_l$ must have at least two elements -- $r_l^{(i)} \in \{ r_l, r_l+1 \}$.  Hubbard
\cite{Hubbard} showed by construction that for all rational $\nu$
there exists a solution in which $S_l= \{ r_l, r_l+1 \}$, and hence that these are ground states.

Before outlining the details of this construction, let us show why maximally compact sets $S_l$ correspond to minimal energy
solutions.   For convex potentials $V(r)$
\ba \label{Def:Convex}
V(r_l +x+y) + V(r_l-x ) >  V(r_l)+ V(r_l +y)
\ea
In other words, given 2 pairs of points with the same average, the
average of any convex function $V$ will be smaller over the pair which
is closest together.  This is simply a reiteration of the familiar
definition of convexity for continuous functions, stating that the line
joining any 2 points on the graph of $V(x)$ lies above the graph.

Now, given two distinct $S_l$ with the same average $\bar{r_l}$,
the convexity of the potential (\ref{Def:Convex}) ensures that the set
with the narrowest distribution of its elements will have the minimum
energy.  More precisely, let $S^*_l$ be the maximally compact set at
level $l$, containing either one or 2 elements, and let
\be
\sum_{r_l \in S_l^*} r_l = M r_l + N (r_l+1) \ \ \ .
\ee
 Any other set $S_l$
with the same mean must by definition contain at least one element
$r_l^{(1)}$ such that either $r_l^{(1)}>r_l +1 $ or $r_l^{(1)} <
r_l$.  (If there is only one element in $S_l^*$ then of course both
limits are given by  $r_l$).  Let us assume that the former case holds, and that
there exists an $r_l ^{(1)} = r_l + 1 +x$ for some integer $x$.  As
both sets have the same average, this means that
\be
\sum_{r_l \in S_l, r_l \neq r_l ^{(1)} } r_l = M r_l + (N-1) (r_l+1) -
x
\ee
In other words, the mean of the remaining elements must be shifted
correspondingly to the left.  Picking one of the shifted elements
$r_l^{(2)} \leq r_l$ (where strict inequality holds if $S^*_l$
contains only $r_l$), create $S_l ^{(1)}$ by shifting $r_l^{(1)} \rightarrow
r_l^{(1)} -1$, $r_l^{(2)} \rightarrow r_l^{(2)} +1$ and leaving all
other elements of $S_l$ unchanged.  Then
\be
E(S_l) -E(S^{(1)}_l)  = V(r_l^{(1)} ) + V(r_l^{(2)}) - V(r_l^{(1)} -1)
-V(r_l^{(2)} +1 >0
\ee
where the inequality follows from the convexity of the potential,
since for $r_l^{(1)} > r_l^{(2)}$ the elements of $S_l^{(1)}$ are more
narrowly distributed about the mean than the elements of $S_l$.
This construction can be repeated so long as some $r_l^{(1)} > r_l+1$
exists -- in other words, so long as $S_l \neq S_l^*$; at each step in
the construction the new configuration has lower energy.  Since this
holds for all $l$, the optimal solution is characterized by $S \equiv
\{ S_l\}_l$ with each $S_l$ maximally compact.

\subsubsection{Constructing the CGS}  \label{HubbardSect}

Hubbard \cite{Hubbard} gives a construction which, for any rational
filling $p/q$, constructs a configuration of particles on the lattice
for which every set of $l^{th}$ neighbor distances is maximally
compact.  A thorough discussion of this solution is given in
\cite{Burkov}; here we present an overview.

Hubbard's solutions are
periodic, and can be expressed in the form:
\be
\overline{n_1 n_2  ...n_j}
\ee
where $n_i$ represents the spacing between the $i^{th}$ and
$(i+1)^{st}$ occupied sites.
Given a continued fraction
$\{ n_0, n_1, ... , n_k \}$, let
\be
\frac{p_i}{q_i} = 1/(n_0 + 1/(n_1+ ... +1/n_i))
\ee
be the numerator and denominator of the related continued fraction $
\{ n_0, n_1, ... n_i \}, i<k$. Now consider
\ba \label{Hub1}
X_0 &=& n_0 \n
Y_0 &=&n_0+1
\ea
with
\ba \label{Hub2}
X_i &=& (X_{i-1})^{n_i-1} Y_{i-1} \n
Y_i &=& (X_{i-1})^{ n_i} Y_{i-1}
\ea
for all $i>1$.  The Hubbard solution is given by
$\overline{X_k}$.  A few examples of these CGS states and their solitons are given in Table
\ref{TheStates}.

\begin{table} \label{TheStates}
\begin{center}
\begin{tabular}{|c|c|c|c|}
\hline
Filling & CGS & particle soliton & hole soliton \\
\hline
$1/q$ & $\overline{q}$ & $q -1$ & $ q+1$ \\
$2/5$ & $\overline{23}$ & $ 22$ & $33$ \\
$3/7$ & $\overline{2 2 3}$ &$ 2 2 2$& $ 323$ \\
$ 5/12$ & $\overline{2 3 2 2 3} $ & $2 2 3 2 2$ & $ 32323$ \\
\hline
\end{tabular}
\end{center}
\caption[Commensurate states]{Occupancy patterns in a few CGS}
\end{table}

One can show by induction that this solution has the
correct filling fraction. The  numerator and denominator of the full continued fraction
can be expressed in terms of numerators and denominators of shorter continued
fractions, via \cite{Khinchin}
\ba \label{Fills}
p_k &=& n_k p_{k-1} + p_{k-2} \n
q_k &=& n_k q_{k-1} + q_{k-2}
\ea
Using (\ref{Hub1}), we have $\overline{X_0} =
\overline{n_0}$ which contains one particle per $n_0$ sites;
$\bar{Y_0} = \overline{n_0+1}$ contains one particle per $n_0+1$
sites.  Equation (\ref{Hub2}) states that if $X_{i-1}$ contains
$p_{i-1}$ particles in
$q_{i-1}$ lattice sites, and $Y_{i-1}$ contains $p_{i-1}+p_{i-2}$ particles in
$q_{i-1} + q_{i-2}$ lattice sites (with $p_{-1}=0$, $q_{-1} = 1$),
then $X_i$ contains $ n_i p_{i-1} + p_{i-2} = p_{i}$ particles in $
n_i q_{i-1} + q_{i-2} = q_{i}$
lattice sites.  $Y_i$ contains $(n_i+1) p_{i-1} + p_{i-2} = p_{i} +p_{i-1}$ particles
in $(n_i+1) q_{i-1} + q_{i-2}= q_i +q_{i-1}$ sites.   Thus at every
step in the process the solution $\overline{X_i}$ has filling fraction
$\frac{p_i}{q_i}$ corresponding to the truncation of the continued
fraction expansion at $n_i$.

Further, it is easy to convince oneself by inspection that these
solutions do indeed satisfy Hubbard's criterion that $S_l = \{ r_l, r_l+1\} $\footnote{ A proof of
this fact can
be found in \cite{Burkov}}.

\subsection{The Devil's staircase}

We have seen how Hubbard's construction gives the ground states of any infinite-ranged convex potential at rational filling fractions.  We now address the issue of these states' stability.
Bak and Bruinsma \cite{Bak} showed that the range of $\mu$ over which each CGS is stable is given by
 \be \label{BakInt}
 \sum_{n=1}^\infty \frac{n q}{(n q+1)^3} +\frac{n q}{(n q-1)^3} - \frac{2 n q}{(n q)^3} \ \ \ .
 \ee
At a filling fraction of $\frac{p}{q}$ the result is independent of $p$, and falls off sharply as a function of $q$.  At $t=0$ these
intervals cover the entire range of $\mu$ pertinent to fillings less
than unity, giving the devil's staircase structure.

To understand Eq. (\ref{BakInt}, one must consider the effect of adding or removing a single particle from the CGS.  We shall see that at filling fraction $\frac{p}{q}$ this produces $q$ fractionally charged solitons of charge $1/q$, each of which distorts the periodic
ground state by altering the distance between one pair of adjacent particles by 1.
(For every commensurate state, there is a unique distortion of this type which minimizes
the potential energy).
 This results in a {\it
 q-soliton state} (qSS).

\subsection{Structure of the q-Soliton State}

To understand why charge fractionalizes, consider the possible
$S_p$ after one charge is added or removed from a chain of length
$L$ containing $N$ particles. At rational filling $N/L = p/q$, for
$l\neq 0 $ mod $p$, we have
 \be
 r_{l_{ij}} \in \{ r_l, r_l +1 \}
\ee 
for all $l^{th}$ neighbors $i$ and $j$ in the HUP CGS. For any
value of $l$, the CGS must satisfy 
 \ba 
N_{r_l} +N_{r_{l} +1} &=&
N \n N_{r_l} r_l +N_{r_{l}+1}(r_{l}+1) &=& l L
 \ea
where
$N_{r_l}, N_{r_{l}+1}$ are the number of $l^{th}$ neighbor pairs
separated by distance $r_l$ and $r_{l}+1$, respectively. After
adding or removing a particle, the most energetically favorable
lattice configuration requires $S_l$ be maximally compact for every $l$. For
$l< p$, this requires that the set of possible radii $\{ r_l,
r_l+1\}$ remain unchanged; for $l=p$, however, we must now assume
$S_p =\{q, q \pm 1 \}$.  These new $p^{th}$ neighbor separations
constitute the solitons of the qSS.  After adding a particle, 
\ba
 N'_{r_l} + N'_{r_{l}+1} &=&  N+1 \n
N'_{r_l} r_l +N'_{r_{l}+1}(r_{l}+1) &=& l L
\ea
giving the net change in the distribution of $l^{th}$ neighbor distances:
\ba \label{counts}
N'_{r_l} -N_{r_l} &=& \pm (r_{l}+ 1) \n
N_{r_{l+1}} -N'_{r_{l}+1} &=&  \pm r_{l}
\ea
where $+, -$ correspond to adding or removing a particle,
respectively.  Substituting $r_p =q$ (one charge
removed) or $r_p =q-1$ (one charge added) into (\ref{counts}) shows that the
qSS contains exactly $q$
solitons.  Thus each soliton has a positive (or negative, in the
case of holes) charge of $1/q$ with respect to the parent lattice.

In a finite system, the qSS is also a HUP state with rational filling,
obtained by distributing the $q$ solitons as evenly as possible on the
chain in order to satisfy Hubbard's criterion at all $l$.  The denominator of such a state will be of the order of the
number of lattice sites.  It
is useful to conceptualize filling fractions whose denominators are
comparable to the system size as being
qSS states of a related CGS of smaller denominator.

Energetically speaking, in a finite system one must
also account for the repulsion between solitons; for most of what follows, we
will drop these terms and consider the infinite volume limit in which
solitons are infinitely far
apart.  In this limit the HUP qSS
consists of $q$ free solitons in the `background' lattice of the
parent HUP state.

\subsection{Energetics of the qSS}

We now turn to the question of the energetics: over what range of
chemical potential is each CGS solution stable against the
formation of solitons? Eq. (\ref{counts}) above shows that the
energy costs of adding and removing particles are, respectively,
\ba \label{Bak} 
E_+ &=& -\mu + \sum_{l \neq 0 (\mbox{mod }p)
}\left[ (r_l+1) V(r_l) - r_l V(r_{l}+1) \right] \n &&+\sum_{l = n q} nq
V(nq-1) -(nq-1)V(nq) + ...  \n 
E_- &=& \mu + \sum_{l \neq 0
(\mbox{mod } p)}\left [ r_l V (r_l+1) - (r_l+1) V(r_l) \right] \n
&&+\sum_{l = n q} nq V(nq+1) -(nq+1)V(nq) + ... 
\ea 
where we ignore
soliton repulsion terms present in a finite system.  Setting $E_\pm =0$ in Eq. \ref{Bak}, 
we obtain the values $\mu_L(p/q) $ and $\mu_R(p/q)$ of the chemical potential at the 
left and right extremities of the plateau.  The width of the plateau (Eq. \ref{BakInt})
is given by the difference between these.  The sum over $l \neq 0 (mod p)$ 
contributes equal amounts to $\mu_L$ and $\mu_R$, 
since $l^{th}$ neighbors separated by $r_l$ and
$r_{l} +1$ both exist in the initial solution, and creating solitons 
simply adjusts their relative frequencies.  However, the solitons replace some $r_p$ by $r_{p} \pm 1$ -- giving different contributions to particle and hole like solitons.
This gives the width of the plateau quoted in Eq. (\ref{BakInt}): 
\ba \label{Bak2}
\mu_R(p/q) - \mu_L(p/q) =\sum_n  nq \left [V(nq-1)+ V(nq+1) -2 V(nq) \right ]
\ea 
which, by the assumption of convexity, is strictly positive.
Since $r_{np}= n q$, the range of stability depends only on $q$.

On the infinite-length chain, the function $\nu(\mu)$ has a complete
Devil's staircase structure.  This means firstly that the function is
monotonic and
contains no finite jump discontinuities \cite{Burkov}.  Secondly, the
set of all such intervals for rational fillings $p/q$ with $q>1$
completely covers the interval $0 \leq \mu \leq E_- (\nu=1)$, and in
the infinite chain limit all ground states are periodic HUP states at
rational fillings.

\subsection{ Proof of devil's staircase structure}

Here we examine a few of the details required to show the devil's staircase structure.
The fact that $\nu(\mu)$ has no jump discontinuities essentially follows from the fact that between 
any pair of rationals on the real line, there is another rational.
We thus focus on the second claim -- that all ground states between $\nu =0$ and $\nu =1$ are commensurate (except possibly on a set of measure $0$).  To do so, it suffices to show that the intervals of stability are disjoint, and that the sum of their lengths is the length of the relevant interval in $\mu$.

To show that the ranges of
$\mu$ over which each rational filling is stable must be disjoint, we first 
point out that as the potential $V$ is a convex function of the inter-particle spacing, 
the energy $E$ is a convex function of the filling fraction.
Indeed, were this not so, it would be energetically favorable at some 
filling fraction to break a system of length $L$ at filling $\frac{p}{q}$ 
into(say) two subsystems of filling $\frac{p}{q} + x$ and $\frac{p}{q} -x$
 -- which, as we know from Hubbard's solution, does not occur.  
This is because if $\epsilon(\rho)$ is the energy per unit length of the 
state at filling $\rho$, then:
\be
2 \epsilon(\frac{p}{q} ) < \epsilon(\frac{p}{q} + x) + \epsilon(\frac{p}{q} -x)
\ee
and the energy is a convex function of filling fraction.

For any $p'/q' < p/q$, choose
$l$ such that $p'/q'= p/q- (l+1)/L$, so that the state 
at filling $(N-l)/L$ contains
$\frac{p'}{q'} L +1$ particles.  The upper boundary of stability of the state $p'/q'$ 
then occurs at the chemical potential where the energy of {\em this} state is 0.
Let $E(N/L)$ denote the lattice contribution to the energy of the state at 
filling $N /L$, such that the total energy of this state is $\mu N + E(N/L)$.
We have:  
\ba
\mu_R\left(\frac{p'}{q'}\right) &=& E\left(\frac{N-l}{L} \right) - E\left(\frac{N-l-1}{L} \right)  \n
\mu_L\left(\frac{p}{q}\right) &=& E\left(\frac{N}{L}\right) - E \left( \frac{N-1}{L }\right) \n
\mu_L\left(\frac{p}{q}\right) - \mu_R\left(\frac{p'}{q'} \right)&=&  E\left(\frac{N}{L}\right) + E\left(\frac{N-l-1}{L} \right) -  E \left( \frac{N-1}{L }\right)- E\left(\frac{N-l}{L} \right)  \n
\ea
which is positive-definite by the convexity of $E$ as a function of filling.
Similarly, for $\frac{p'}{q'}> \frac{p}{q}$, we choose $p'/q'= p/q+ (l+1)/L$
\ba
\mu_L \left(\frac{p'}{q'} \right) - \mu_R \left(\frac{p}{q} \right ) 
= 
 E\left(\frac{N+l+1}{L}\right) + E\left(\frac{N}{L} \right) -  E \left( \frac{N+1}{L }\right)- E\left(\frac{N+l}{L} \right)  \n
\ea
which is again non-negative by convexity.
Thus the intervals of stability of different rational
fillings do not overlap.  

It remains to show that the sum of the intervals of stability of all
rational fillings completely covers the range $0 \leq \mu \leq
\mu(1)_L$.  According to (\ref{Bak}), the chemical potential $\mu(1)_L$ at which the
$\nu=1$ state becomes unstable is:
\ba
\mu(1)_L &=& \sum_{n=1}^\infty (n+1) V(n) - n V(n+1)  \n
&=& 2 \sum_{n=1}^\infty V(n)
\ea
To compute $\sum_{x \in Q} ( \mu_R (x)- \mu_L(x)$, we must add together the 
length of the interval for all possible denominators $q$, multiplied by the 
number of reduced fractions with this denominator.  
The function which gives this multiplicity is 
the Totient function $\phi(q)$, which counts the number of integers less than
$q$ which are relatively prime to $q$. Hence the combined length of all intervals is
\ba
I &\equiv & \sum_{p/q}\left [\mu_R \left(\frac{p}{q}\right) - \mu_L \left (\frac{p}{q}\right) \right ] \n
&= &\sum_{q=2}^\infty  \phi (q) \sum_n  nq \left
  [V(nq-1)+ V(nq+1) -2 V(nq) \right ]
\ea
 Collecting the contributions
to $V(r)$ for each $r$ gives
\be
I = \sum_r \left [ \sum_{q | r+1} \phi (q) (r+1) + \sum_{q| r-1} \phi(q) (r-1) - 2
\sum_{q|r} \phi(q) r \right ] V(r) \ \ \ .
\ee
 Using the identity\footnote{Found on Wikipedia!!}
\be
\sum_{q|r} \phi(q) =r
\ee
gives
\ba
I &=&\sum_r \left [ (r+1)^2 +(r-1)^2 - 2 r^2 \right] V(r)\n
&=& 2 \sum_r V(r)
\ea
which is precisely the boundary of stability of the $\nu=1$ state,
establishing the desired result.

\section{ Away from the classical limit: Mott-Hubbard transitions in the strong coupling
expansion}  \label{PertSect}

The physics of the Mott transition in one dimension has been studied in the context of classical phase boundaries, as the commensurate-incommensurate transition, as well as in quantum mechanics via the Hubbard and extended Hubbard models.  Basically, as quantum or thermal fluctuations in particles' positions increase, commensurate order is destroyed by a condensation of solitonic defects.  In the quantum mechanical case, these defects cost potential energy, but are favored by kinetic terms, driving a transition at some finite hopping on the lattice.  This physical mechanism is also responsible for a commensurate- incommensurate transition in systems with infinite-ranged convex interactions.  However, the fractal structure of the classical ground states leads to a phase portrait considerably more complex than in cases studied previously.  Here we will examine the structure of this phase portrait.

The
qualitative behavior of our system in the convex regime is reminiscent of
the Bose-Hubbard model, with commensurate Mott lobes ceding to
superfluid states as $t$ increases.  It is convenient to treat a state with large $q$ as a state with smaller $q$ at a nearby filling in which a
crystal of dilute solitons has formed. Hopping tends to liquify
dilute crystals of solitons: at large separation the inter-soliton
repulsion is smaller than the kinetic energy gained from
delocalization.  (The latter grows as $~ 1/r^2$, and the 
former as $\frac{1}{r^3}$, at large separations).  
The delocalized solitons destroy long-ranged
spatial order, creating a Luttinger liquid with full translational
symmetry. Hence, as $t$ increases, the system undergoes a transition
from the Mott insulating CGS to a Luttinger liquid state, with
larger $q$ states liquifying at smaller $t$.

\subsection{ Strong Coupling Expansion }  \label{tpert}

To find the position of
the phase boundary, we generalize the method of
Ref.~\cite{FreericksMonien} and compare the energies of the CGS and
its adjacent qSS to third order in $t$  using standard
time-independent perturbation theory. This approach assumes that the
phase transition is continuous, so that for a given $t>0$, values of
$\mu$ for which $E_{qSS}(\mu) = E_{CGS}(\mu)$ constitute the phase
boundary. This assumption is well-founded, since the soliton
repulsion ensures that the energy cost of creating multiple solitons
is larger that of a single soliton, thus favoring a second-order
transition.

In a finite system one must account for the repulsion between
solitons; here we drop these terms and consider the infinite
volume limit. In the limit that the solitons are sufficiently well
separated that we may neglect their interactions, the qSS is
highly degenerate and can be expressed in terms of a band of solitonic
momentum eigenstates. Here we consider only the
bottom of the band, which lies at zero momentum.

To find the transition, we compare the energies
(calculated up to third order in $t$) of
the CGS (which is the $t=0$ ground state) and the qSS.
The calculation of the energies is carried out using standard
time-independent third order perturbation theory.  The energy
corrections are calculated in terms of the $t=0$ ground and excited
states of the CGS and qSS.

\subsubsection{Perturbation theory in the CGS}

To calculate these energy corrections, we must consider excitations about the classical
CGS generated by perturbative hopping, and their matrix elements with the unperturbed state.
 The zeroth order CGS is given by
the HUP solution $|\p(CGS) \rangle$, which is
non-degenerate (up to global translations). Hopping creates excited states of the form:
\be \label{CGSex}
|\psi_{ex}^{(0)} (x) \rangle = b^\dag_x b_{x+1} |\p(CGS) \rangle
\ee
At filling $\frac{p}{q}$ there are exactly $2p$
distinct such hoppings which must be considered: one in each direction
for each occupied site in the HUP unit cell.

To calculate the energy corrections, we must calculate matrix elements of
the ground and excited states with $H_1$, and the energy differences
$\delta E_i$ between the corresponding classical ground and excited
states.  Note that the first order wave function contains only terms which
can be transformed into the classical ground state by only one hopping --
in other words, only the excited states of the form (\ref{CGSex}) and
(\ref{qEx}) have non-zero contributions.
The details of the perturbative approach are outlined in Sect.  \ref{PertApp};
here we will only state the results.

The first and third order corrections to the CGS energy are zero,
because the CGS is non-degenerate and hence any odd number of hoppings
produces a state orthogonal to the ground state.
At $\nu = p/q$, the second-order correction is given by
 \ba
  E^{(2)}_{CGS} &=& - 2 \frac{N t^2}{p} \sum_{i=1}^p \frac{1}{\Delta E_i}
 \ea
where $\Delta E_i = E_i^{(0)}-E_0 ^{(0)}$ is the difference in
potential energies between the ground and the excited state formed
by hopping from the $i^{th}$ occupied site in the ground state
configuration.  As the ground state is periodic, it suffices to
calculate these energies for the $p$ distinct particles in the
repeated pattern.

\subsubsection{Perturbation theory in the qSS}

We now repeat this analysis for incommensurate fillings, which
have solitonic ground states. The qSS is given at zeroth order by
the HUP qSS, consisting of $q$ HUP solitons sufficiently far apart
that soliton-soliton interactions can be ignored. The energy of
the qSS is then simply $q$ times the energy of a single soliton on
an infinite lattice. The qSS ground states are highly degenerate
in this limit, since all translates of each soliton have equal
energy.  We therefore work in the basis in which the hopping term
is diagonal, namely: \be |\psi_{qSS}^{(0)} (k) \rangle = \sum_x
e^{ik x } | \psi_{qSS} (x) \rangle \ee where the sum runs over
occupied lattice sites, and $|\psi_{qSS}(x)\rangle$ is the state
containing one soliton beginning at lattice site $x$.  The soliton
hops by $q$ sites when a single boson on one of its edges is
hopped by one site; hence the position $x$ is an integer multiple
of $qa$.  The perturbation theory is essentially the same 
for qSS states containing solitons and anti-solitons; 
functionally the difference 
between these is in the energy gaps to the local excitations.

Local excitations, or defects, of the qSS are again generated by hopping 
one particle away from its preferred position.  
These can be expressed in the
form: 
\be \label{qEx} |\p_r(k) \rangle = \sum_x e^{i k x }  b(x+r)
b^\dag(x+r\pm 1) |\p^{(0)}_{qSS} (k) \rangle 
\ee 
The state
$\p_r(k)$ describes correlated propagation of a soliton and a defect
$r$ lattice sites away\footnote{ In the following the word
soliton applies strictly to the HUP solitons described in section
(\ref{HubbardSect}); lattice distortions due to other hoppings we
will call defects.  The two are indeed different as solitons carry global 
topological charge, while defects -- which constitute local re-arrangements 
of the charge-- do not.}.  Note that for
each $r$ the defect has 2 possible orientations, depending on
whether the hopping has been towards or away from the defect.

Computing the resulting matrix elements gives the energy corrections:
 \ba
E^{(1)}_{qSS} &=& -2 qt  \cos (kq a )  \n
 E^{(2)}_{qSS} &=& 2 q \cos(2 k q a) \frac{t^2}{\Delta E_{r_1, -1}}
 - q \sum_{i=1}^{N/q} \sum_{\alpha=\pm 1} \frac{t^2}{\Delta
  E_{r_i, \alpha}}\n
 E^{(3)}_{qSS} &=&- 2 q t^3 \cos(k q a)  \left [  \frac{\cos (2 k q
a)}{\Delta E_{r_{1, -1}} \Delta E_{r_{1, -1}+q}} - \frac{\cos (2 k
q a)}{(\Delta E_{r_{1, -1}})^2}  \right ]\n
 & & -2q t^3 \cos(kqa)
\sum_{i=1}^{N/2} \left [\frac{1}{(\Delta E_{r_i})^2} -
\frac{1}{\Delta E_{r_i} \Delta E_{r_{i+1}}} \right ]
 \ea
Here $k$ is the soliton momentum, $a$ is the lattice constant, and
$q$ is the denominator of the commensurate filling fraction, which
appears here because one hopping displaces the soliton by $q$
lattice spacings.  The subscripts on $E_{r_i, \pm1}$ indicate the
distance between particle $i$ and the soliton, and the direction
of the hopping relative to the soliton. The special distance
$r_{1} $ describes an excited state in which an anti-soliton is
sandwiched between 2 solitons (for a solitonic qSS), or vice versa 
(for a qSS with anti-solitons).  These states
contribute extra terms to the energy corrections of the qSS
because of the ambiguity as to which soliton is associated with
the ground state qSS.

\begin{figure}[htp]
\begin{center}
\includegraphics[totalheight=10cm]{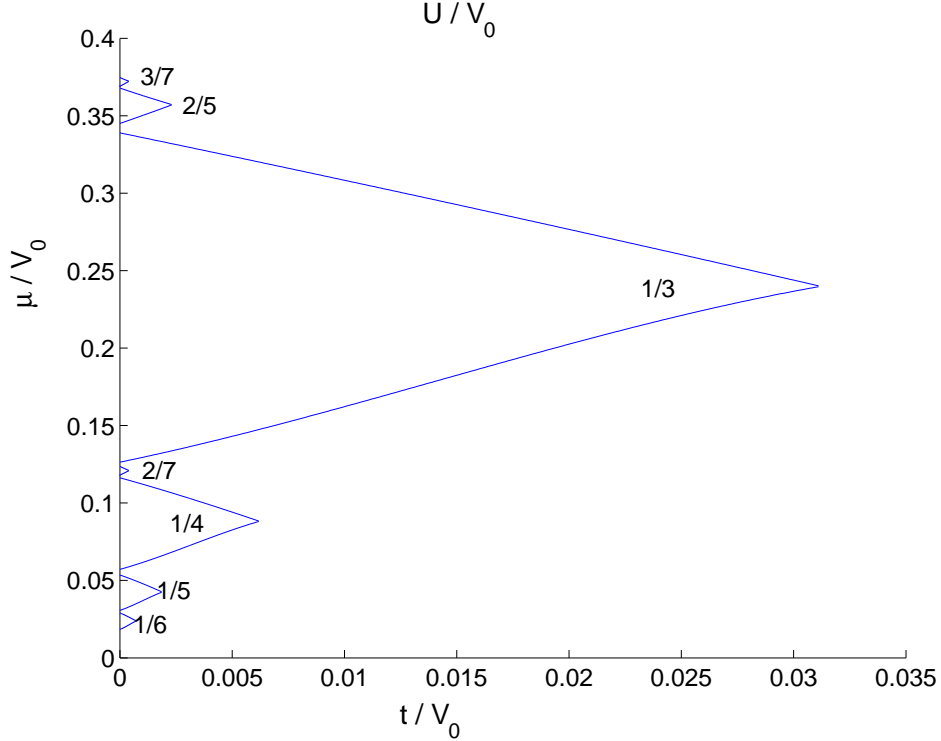}
\end{center}
\caption[Perturbative calculation of the Mott to SF
phase boundary]{\label{Hoppings} Perturbative calculation of the Mott to SF
phase boundary in the $(t, \mu)$ plane, shown here for $U=20$.  Here
the strength of all couplings is measured relative to that of the
dipolar interaction strength $V_0$.  Each lobe encloses a Mott
insulating region in which the filling is fixed; the region outside
the lobes is a superfluid of solitons.  The $1/3, 1/5, 1/6, 2/7,
2/7$ and $3/7$ -filled lobes are shown here.  Every commensurate
state of the complete Devil's staircase has a Mott lobe, but the
range of hoppings over which a state exists falls off sharply with
its denominator. }
\end{figure}

Fig.~\ref{Hoppings} shows the results of the perturbative
calculation for selected filling fractions.  The $t=0$ axis
corresponds to the classical limit, in which the CGS states
comprise a complete devil's staircase: every value of $\mu$
corresponds to a rationally filled ground state, except for a set
of measure 0.   The figure shows the resulting Mott lobes: inside
each lobe the CGS is stable and the system is in a Mott insulating
state.  The Mott gap vanishes on the boundary of the lobe;
outside of this region solitons proliferate and the system is in a
Luttinger liquid phase.

For any $t>0$, only a finite number of insulating states exist; the
rest are liquid states with a superfluid of condensed solitons. The
function $\nu (\mu)$ is no longer a devil's staircase, but rather a
piecewise smooth function, with plateaux of constant density
separated by liquid phases whose density varies continuously with
$\mu$.   The size of the commensurate region decreases sharply with
$q$: states of higher $q$ have both smaller ranges of stability in
the classical limit, and larger energy corrections relative to the
CGS, so that the volume of the corresponding Mott lobe scales
approximately as $1/q^5$.\footnote{ This scaling uses a first-order
in $t$ approximation to the boundaries of the Mott lobe.}  The total
volume occupied by liquid states can be estimated from the
first-order approximation to the Mott lobe boundaries; we find that
for small but fixed $t$ the volume of the liquid region scales as
approximately $t^{2/5}$.  (Details of the
calculation are given in Sect.  \ref{QScale}).
At sufficiently large $t$ we expect all
insulating states to be unstable, and the particle density to vary
smoothly with $\mu$.

\subsection{Bosonization treatment of the phase transitions}

The qualitative nature of the Mott transition can be deduced from
existing knowledge of 1D commensurate-incommensurate phase
transitions, which we summarize here. We approach the phase transition 
from the Luttinger liquid side, where bosonization can be used to
treat the kinetic term and dipolar interactions exactly.   The lattice
potential can be added perturbatively; when the coupling of the 
density to the lattice becomes relevant, this signals the 
Mott transition.  

In bosonized form, our
system is described by a Hamiltonian that is a sum of
a Luttinger piece, accounting for the kinetic term and dipolar
interactions~\footnote{Though the potential is infinite ranged, it
falls off quickly enough that the qualitative description is
identical to that for short-ranged interactions~\cite{Giamarchi}.  This 
is true for any long-ranged potential with a finite Fourier transform 
at $0$ momentum.},
and a sine-Gordon term which emulates the lattice in the continuum
limit.  We also include a parameter $\delta$, which parametrizes the 
deviation from commensurate filling.
Close to filling $\frac{p}{q}$, the bosonized Hamiltonian is thus:
\be \label{Luttinger}
H = \frac{1}{2 \pi} \int dx \left [ u K (\pi \Pi(x))^2 + \frac{u}{K}
  (\bigtriangledown \phi(x))^2  + g \cos (2 q \phi(x) - \delta x) \right ]
\ee 
Here $\phi$ is related to the density of the physical bosons
by 
 \be \label{phidens} 
\rho(x) = \left[ \rho_0 - \frac{1}{\pi}
\bigtriangledown \phi(x) \right ] \sum_n e^{i 2 n ( \pi \rho_0 x-
\phi(x)) } 
 \ee 
and $\Pi$ is the momentum conjugate to $\phi$.  $u$
and $K$ are the usual Luttinger parameters, expressible in terms
of the hopping and interaction terms of the original system. The
sine-Gordon term adds a periodic lattice potential 
\be \label{Eq.LatticeCouple}
 \int d x
\rho(x)  V \cos(\frac{2 \pi }{a} x) 
 \ee 
to the model; 
$g$ parameterizes the strength of the coupling to this lattice.

Near commensurate densities $\rho_0 = \frac{1}{qa} + \delta$, the expression 
(\ref{phidens}) for the density shows that the slowly varying modes are 
precisely those for which
\ba \label{Res} 
n = q 
\ea 
where $n$ is the resonant
harmonic from the density (\ref{phidens}).  
The resulting expression for the coupling (\ref{Eq.LatticeCouple}) 
to the lattice is
\be \label{Eq.Dens} 
V \int dx \cos(2 q \phi(x)- \delta) \ \ \ .
 \ee 
For densities $\frac{p}{q}$ we obtain the same expression, 
since the state at filling $\frac{p}{q}$ consists of several 
charge density wave instabilities at wave vector $q$, whose 
relative positions also become pinned inside the Mott region. 

The Hamiltonian (\ref{Luttinger}) is well studied in both the context of the
Mott transition \cite{Giamarchi} and the Frenkel-Kontorowa model of
surface interfaces \cite{Schulz}.  The upper and lower Mott lobes
join in a cusp which is not accurately described by the perturbation
theory-- since `small' hopping implies that the ratio $t/ \delta
E$ of the hopping relative to the energy gap to the solitonic
states must be small, the range of $t$ over which the perturbative
treatment is valid decreases with $q$ and never encompasses this
point.

 The bosonized treatment reveals that crossing the edge of
the Mott-Hubbard lobe induces one of two different types of phase
transitions. At the cusp joining the upper and lower Mott lobes, a
constant-density phase transition, with $\delta =0$ in Eq. (\ref{Eq.Dens}) of the Kosterlitz-Thouless type
occurs. For $\delta \neq 0$ the transition is well described by a simple
two-band model with quasi-particles that are gapped in the
commensurate phase, and with a density increasing as $\sqrt{ \mu -
\mu_c}$ near the transition on the liquid side.

\subsubsection{Derivation of the Luttinger description near the phase transition}
\label{Sect_LuthEm}

To derive these results, the Hamiltonian (\ref{Luttinger}) must be
treated using a perturbation series in $g$.  This is easy to do
for $\delta = 0$. One finds that for $K<K_c$, the cosine term
becomes relevant at low energies, and locks the state into a
crystalline phase commensurate with the lattice.  For $K> K_C$,
so-called domain walls (in our case, lattice solitons and
anti-solitons) proliferate, and the phase becomes a liquid.  In
the renormalized Hamiltonian, this predicts a transition from the
liquid to the solid phase at $t \sim 1/q^4$ at filling $\frac{p}{q}$, consistent with the
perturbative results of Section \ref{tpert}.  At the transition,
$K$ jumps discontinuously from the universal value $K_c = 1/2$ (approaching
from the
liquid state) to $0$ in the gapped Mott phase.  This is the standard signature of a
Kosterlitz-Thouless transition.

A phase transition with $\delta =0$ occurs only for the value of $\mu$ for which the
model is particle-hole symmetric.  For
non-zero $\delta$, the situation is similar, but at constant $\mu$ the renormalization
of the doping flows to strong coupling before the phase transition; hence 
the transition is not accurately described by perturbation theory about the 
Luttinger Hamiltonian.  In this case
the transition is best studied using the Luther-Emery solution-- that
is, re-scaling the problem and
exploiting the fact that it can be mapped onto a system of weakly
interacting fermions (in our case actually the lattice solitons) with an
upper and lower energy band separated by a gap.

To do this, we first scale the fields in Eq. (\ref{Luttinger}) according to $\tilde{\phi} = q \phi + \delta x /2,
\tilde{\Pi} = \Pi/q$.  Then
\be \label{HL}
H = \frac{1}{2 \pi} \int dx \left [ u \tilde{K} (\pi \tilde{\Pi(x)})^2 + \frac{u}{\tilde{K}}
  (\bigtriangledown \tilde{\phi(x)} - \delta x /2 )^2 \right ] + \cos (2
\tilde{\phi(x)})
\ee
where $\tilde{K} = K q^2$.  The C-IC transition occurs at a value of $K
= \frac{1}{2 q^2}$, or in this picture $\tilde{K} = 1/2$.
Here we will drop all $\tilde{ }$, and work in the scaled system.

Second, we translate the Luttinger Hamiltonian (\ref{HL})
back into the language of spinless fermions:
\ba
H &=& \sum_k v_F (c^\dag_{R k} c_{R k} - c^\dag_{L k} c_{L k} ) + \Delta
(c^\dag_{R k} c_{L k} + c^\dag_{L k} c_{R k} )
\n
&&+  g_2( \rho_R(k)
\rho_L(-k) + H.C) + g_4 \left [ \rho_R
(k) \rho_R(-k) + \rho_L(k) \rho_L(-k) \right ]
\ea
This is convenient because
the potential $\cos 2 \phi$ due to commensuration effects has been mapped to
the quadratic term $c^\dag_R c_L + h.c.$.  Thus at $\tilde{K} =1$, along the 
Luther-Emery line where the system is particle-hole symmetric,
the interaction terms $g_2$ and $g_4$ vanish and we have:
\be \label{Thirring}
H_0 = v_F k (c^\dag_{R k} c_{R k} - c^\dag_{L k} c_{L k} ) + \Delta
(c^\dag_R c_L + h.c.)
\ee
This simply describes free massive fermions, of energies $ \epsilon =
\pm \sqrt{ v_F^2 k^2 + \Delta^2}$, with the mass gap $\Delta$ is set by
the coupling of the cosine term.
For $\tilde{K} \neq 1$, Schulz \cite{Schulz} showed that a parametrization
can be chosen such that the $g_4$ term vanishes, and the coupling of the 
remaining interaction is proportional to the doping away from commensuration,
which vanishes at the phase transition.  Thus the non-interacting model 
gives a good description of the physics very near the phase transition.

In practice, the quasi-particles in (\ref{Thirring}) may be identified with
Hubbard solitons (or distortions of the commensurate lattice) which
become gapless and proliferate at the transition.
 In the
solid phase, the solitons are gapped, and do not occur at sufficiently
low temperatures.  At the phase boundary the chemical potential
crosses the bottom of the upper band, and solitons proliferate.  Near
the transition on the liquid side, this predicts
$
d \sim \sqrt{\mu - \mu_c}
$
with $\mu_c$ the chemical potential at the phase transition.

\section{Effects of the Trapping potential}  \label{FiniteTrapSect}

In practice, any experimental
realization of the HUP model will involve a finite sized atom trap,
generally of length not more than a few hundred lattice sites.  In
addition, the trapping potential is not generally flat, but rather is
well approximated as a harmonic potential.  Here we will try to
address the possible effects of this trap on the system.

We can estimate the effects of a harmonic trapping potential
(present in current cold-atom experiments) using the local density
approximation -- this assumes that the trapping potential is
slowly varying enough that it can be simply incorporated into the
chemical potential, resulting in a spatially-dependent $\mu$.
Trajectories along the 1D chain then correspond to cuts in
Fig.~\ref{Hoppings} at fixed $t/V_0$. Thus, at $t=0$ we expect
different commensurate fillings at different points along the
trapped chain, with the most stable states (fillings $1/2$ and
$1/3$) occupying the largest regions within the trap. However,
commensurate states with a period $q$ greater than the lengthscale
over which the trapping potential varies cannot exist in the
trapped system since they violate the local density approximation.
The situation is improved slightly when $t$ is nonzero but small,
where small islands of Mott states at various fillings $p/q$ are
separated by regions of superfluid.  These fluid regions can interpolate 
continuously in density
between the two commensurate states. 
We will discuss the
resulting density profiles for various chemical potentials.  In doing so,
we ignore the CDW correlations of the liquid states, which near filling $p/q$ 
also occur at length scale $q$; hence this treatment is a modest improvement 
of the LDA.

\subsection{$t=0$ physics and the LDA}

The simplest approach to determining the density profile in a finite
harmonic trap is to use an effective Local
Density Approximation (LDA) treatment.  In other words, given an
approximation to the
function $\nu(\mu)$ as calculated by \cite{Bak}, one can generate a
profile of the filling $\nu(x)$ as a function of position, $x$, in the
trap, treating the trapping potential as a spatially varying chemical
potential.   In this approach, we neglect the fact that each filling $p/q$
requires at least $q$ lattice sites to be realized, and that the range
of $x$ over which $\nu= p/q$ may well encompass less than $q$ sites.

\begin{figure}[htp] \label{LDAs}
\begin{center}
\includegraphics[totalheight=5in]{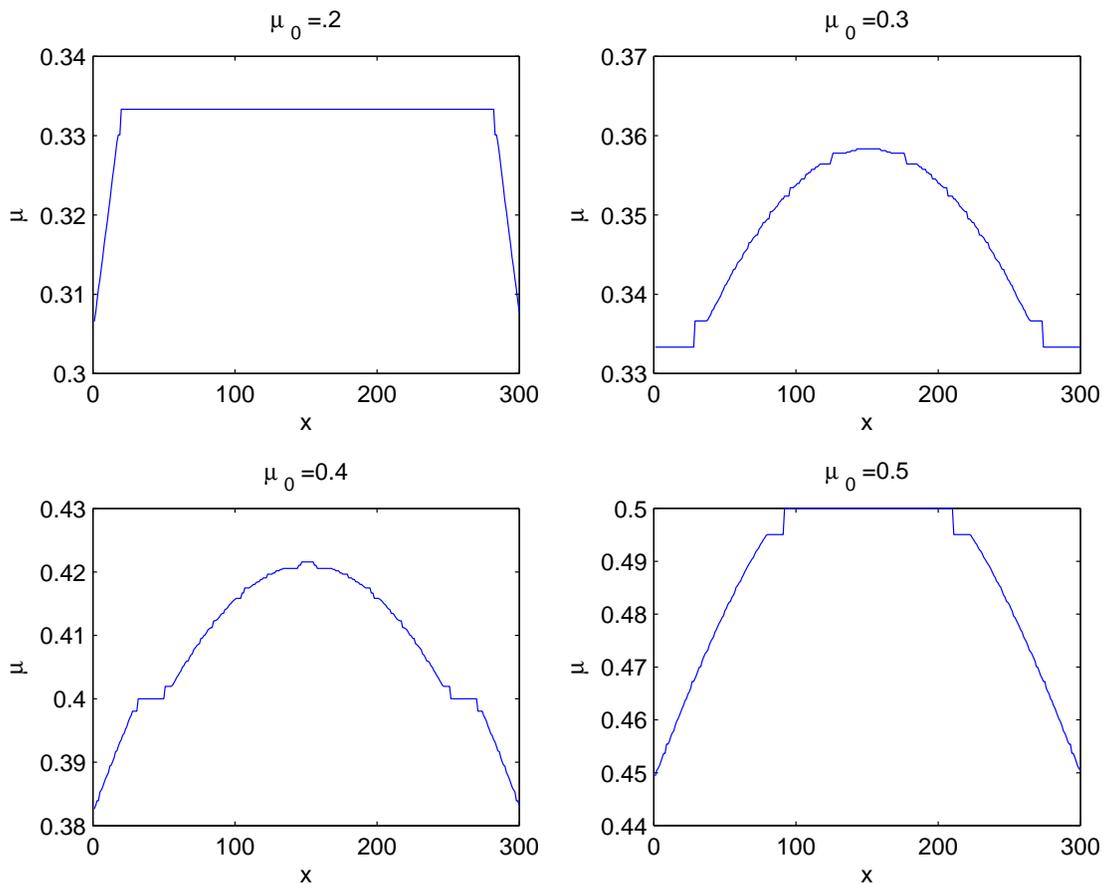}
\end{center}
\caption[Local density approximation for filling profile in a finite trap]{Local density approximation (based on the HUP solution for
  infinite $\mu$) for a trapping potential $V(x) =(x-x_0)^2/(L/4) a^2 -
  \mu_0$ (where $L$ is the total number of sites in the trap) shown
  for various values of $\mu_0$.  We expect the large
  plateaux of small denominator states to remain as features in the
  actual finite-volume solution, but that most smaller plateaux will
  disappear into regions of ambiguous density.}
\end{figure}

Figure \ref{LDAs} shows sample LDA profiles for various values of the
trap minimum $\mu_0$, taken from a numerical computation of $\nu(\mu)$ cut off
at $q=200$.  For some parameter choices, most of the trap lies in a
chemical potential range that is strongly locked at a commensurate
filling with low denominator, and the LDA gives a plausible rendering
of the density profile over much of the trap.  However the figure also
shows clearly that for some regions of the trap the LDA is a very poor
approximation, as it predicts
fillings $\nu =p/q$ with
$q$ much too large to fit into the number of lattice sites over which
the state is stable.

\begin{figure} [htp] \label{TrapsNum}
\begin{center}
\includegraphics[totalheight=5in]{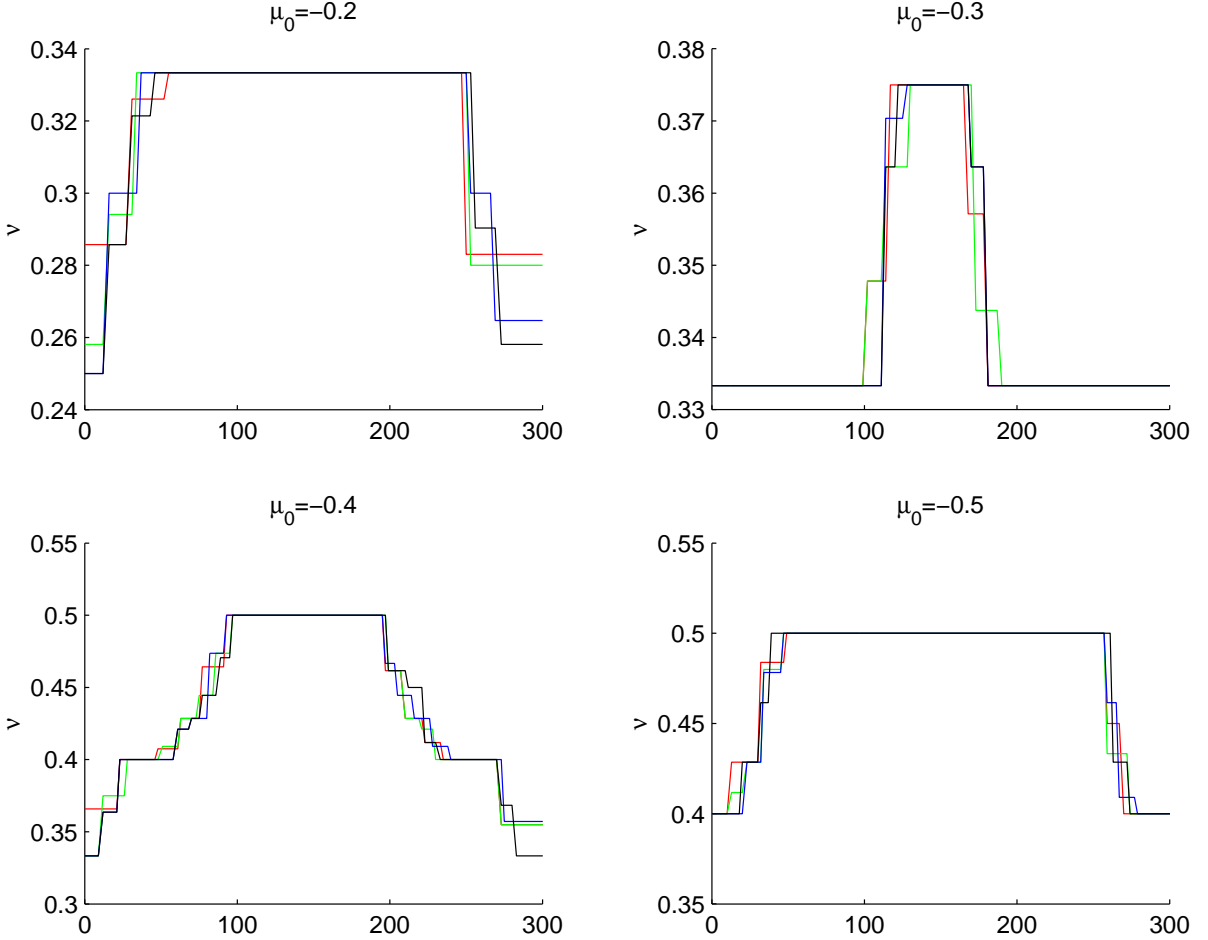}
\end{center}
\caption[Density profiles in a finite trap calculated using simulated annealing]{Density profiles calculated using simulated annealing for
the same potentials as in Figure \ref{LDAs}.  The multiple lines show 
multiple runs of the simulation, indicating that the states at low-denominator 
filling fractions are very robust, while states at higher values of $q$ ($q >4$)
are energetically delicate and require simulations over longer time periods than 
those undertaken here.}
\end{figure}

In the classical limit, another approach is to simulate the density
profile numerically.  We have used simulated annealing to generate
profiles; Figure \ref{TrapsNum} shows numerically
generated profiles for the same trapping potentials as in Figure
\ref{LDAs}.  Note that in regions of transition between different
commensurate states, the exact density is somewhat ambiguous, and can
vary depending on what technique is used to calculate it.

Though neither of the techniques discussed above proves
particularly illuminating in the transition areas between plateaux
of relatively small $q$ states, they illustrate some practical
features of the HUP system in a trap.  First, the spatial
homogeneity of the density depends strongly on the depth of the
trap.  Since the range in $\mu$ over which the lowest $q$ states
($\nu =1/2, 1/3 ...$) are stable is much larger than the ranges of
stability of higher denominator states, a trap with a modest
curvature can be `locked' everywhere at a density of $1/2$.  As
$\mu_0$ is shifted, however, portions of the trap fall outside the
range of half-filling, and since all of the nearby states have
much larger denominators, their ranges of stability will be
correspondingly much smaller.  In a trap of only a few hundred
sites, for the most part these states will not be stable over a
wide enough range to produce a strong scattering peak.  Hence it
is the strongly locked regions which are the most visible
experimentally, and these, fortunately, are well -described both
by LDA and simulated annealing techniques.

\subsection{Spatial Profiles of Atoms in a harmonic trap at finite $t$}

We have seen that the LDA gives a poor description of the expected profiles
for high-denominator filling fractions in a finite trap.  However, in practice
such states would not in any case exist in any currently attainable experiment.
Hence in practice we expect density profiles which consist of phase separated
regions of density-locked Mott states interspersed with compressible fluid
states whose density varies continuously with the trapping potential.

The qualitative properties of the
quasiparticle fluid, at least near the transition, are well described by the
Hamiltonian (\ref{Thirring}).  To describe the density of the Luttinger liquid
regions, we make a quantitative mapping between
the 2-band model (\ref{Thirring}), and
the perturbative description of the quasi-particle fluid valid for small
$t$.

If we ignore the weak interactions between quasi-particles, this effective
description has 2
free parameters: the effective Fermi velocity in the upper band, and the
band gap.  By calculating these parameters using the perturbative hopping
model, a quantitative matching between the Luther-Emery solution described in
Sect. \ref{Sect_LuthEm} and our system can
be achieved.  We will use this model to generate profiles of particle
densities in the trap at finite filling.

The band gap $\Delta$ is given simply by the difference in energies of the
IC states with one extra particle and one extra hole.  At $t=0$ this is
just the devil's staircase pattern calculated by \cite{Bak}; at finite
$t$ it can be extracted from the perturbative calculation of Sect.
\ref{tpert}.

To match the Fermi velocities, we match the true free
soliton Hamiltonian:
\be H^{(t)} = s^\dag (k) \left[ \Delta + \mu +t (1-
\cos (q a k)) \right ] s (k)
\ee
onto the effective Hamiltonian
\be H^{(S)} = s^\dag (k) \sqrt{ (v k)^2 + \Delta^2} s(k)
\ee
where $v$ is
the Fermi velocity in the original 2-band model.

Both $H^{(t)}$ and $H^{(s)}$ are quadratic in $k$ near the bottom of the
upper band.  The effective model describing the dynamics of particles in
the upper band only is obtained in both cases by linearizing the spectrum
about $k_F$ in the upper band.  This gives:
\ba
\tilde{H^{(t)}} &=& s^\dag
(k) \left [\Delta + \mu +t(1-\cos(aq k_F)) + k a q t \sin(a q k_F) \right
] s(k) \n
\tilde{H^{(s)}} &=& v^2 k_F k / \sqrt{(v k_F)^2 + \Delta^2}
\ea
Of course $\mu$ is such that the constant terms in the first equation
cancel, and both energies are linear in $k$.  Hence we can match:
 \be v^2
k_F /\sqrt{(v k_F)^2 + \Delta^2} = q a t \sin(q a K_F)
\ee
To lowest order
in $k_F$ (valid at low quasiparticle densities) this gives:
\be
\label{getV} v^2 = (q a)^2 \Delta t
\ee

We may use this result to calculate the density of solitons in the
IC region near the phase transition.  In particular, if we ignore
quasiparticle interactions, we have \ba N &=& \frac{L}{2 \pi v}
\int _\Delta^\mu \frac{\epsilon}{\sqrt{\epsilon^2-\Delta^2}} d
\epsilon \n &=& \frac{L}{2 \pi v} \sqrt{\mu^2 - \Delta^2} \ea
where in the non-interacting model $\Delta = \mu_c$ is the
chemical potential at the transition, which can be estimated using
the perturbative calculation of the phase diagram. (Interactions
will renormalize the gap in principle, though in our case we
calculate the gap perturbatively, and this result should be
accurate for sufficiently small $ t$.)

Substituting in the value of $v$ obtained in equation (\ref{getV}),
and accounting for the fact that each soliton has a charge density of
$\frac{1}{q}$ relative to the background charge density of the
lattice, we obtain the charge density near the C-IC transition:
\be \label{DensProf}
\rho (q) = \frac{1}{2 \pi q^2 \sqrt{\mu_c t} } \sqrt{ \mu^2 -\mu_c^2}
\ee
This gives the desired expression for the density of quasi-particles
as a function of the parameters of the original HUP Hamiltonian.

\subsection{Density profiles}

\begin{figure} [htp] \label{tProfiles}
\begin{center}
\includegraphics[totalheight=5in]{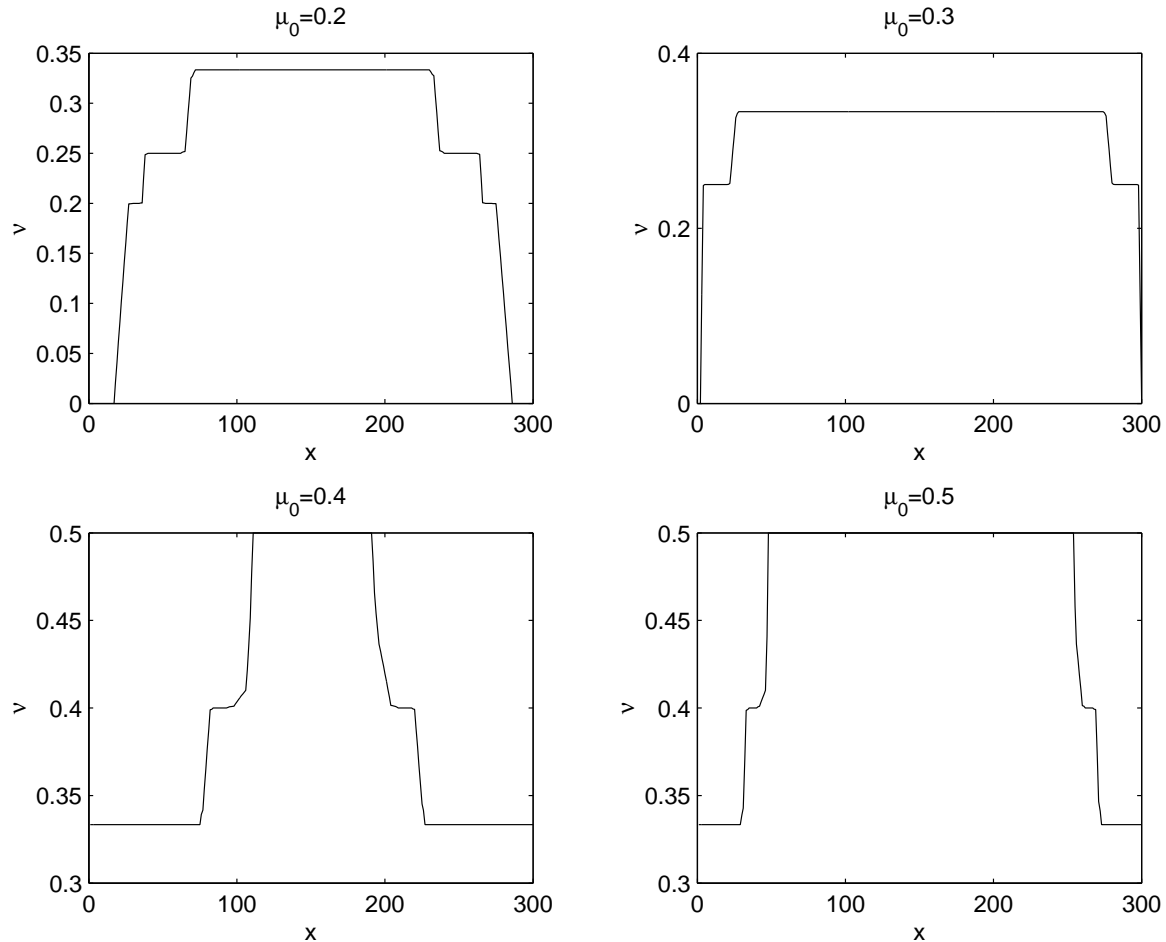}
\end{center}
\caption[Density profiles in a harmonic trap at finite hopping I]{Density profiles in a harmonic trap at $t/V =0.001$.  Stable plateaux
can be seen at $1/2, 2/5, 1/3$, and $1/4$ filling, separated by Luttinger liquid
regions of continuously varying density.}
\end{figure}

\begin{figure} [htp] \label{tProf2}
\begin{center}
\includegraphics[totalheight=5in]{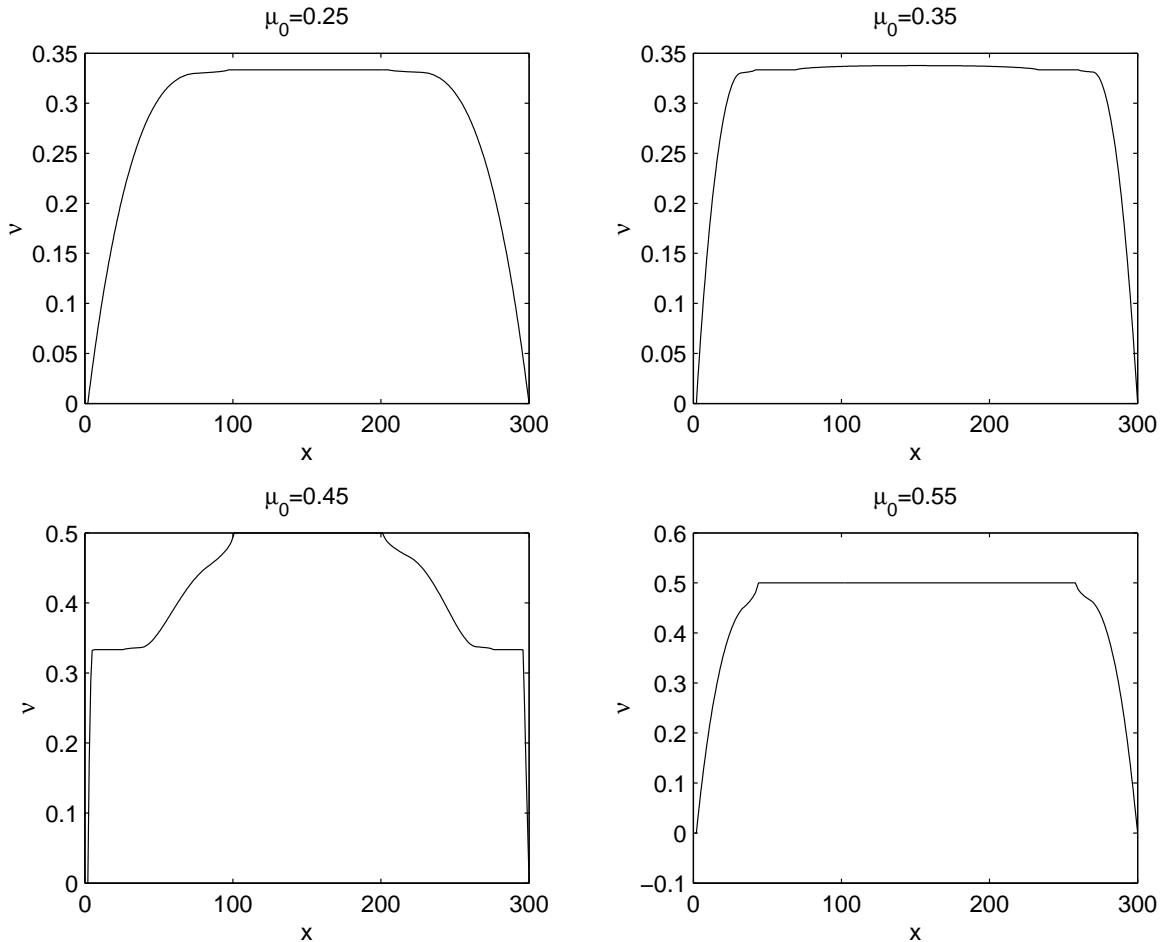}
\end{center}
\caption[Density profiles in a harmonic trap at finite hopping II]{Density profiles in a harmonic trap at the experimentally
realizable value $t/V =0.02 $.  Here only the $1/2$ and $1/3$
filled commensurate states are stable.}
\end{figure}

We now use the results of the previous section to estimate the
density profiles in a realistic trap, by using a finite $t$ LDA.
This scheme first identifies the commensurate regions of the trap;
since for the values of $t$ considered here all but a few very
low-denominator fractions are unstable, the problem of patterns
too short to fit into the allotted number of lattice sites does
not arise.  At the edges of each commensurate region we use
(\ref{DensProf}) to predict the local density profile;  a
polynomial interpolation is used to join the various liquid
regions.

Figure  \ref{tProfiles} shows typical charge density profiles in a
trap $t/V = 0.001$.  Commensurate plateaux at filling fractions $1/2,2/5, 1/3$,
and $1/4$ can be stabilized, depending on the chemical potential at the bottom of the trap.
Between these plateaux we see regions of Luttinger liquid.

Figure \ref{tProf2} shows density profiles for $t/V =0.02 $, the value in
principle attainable by experiments on polar molecules.

Though we do not claim to predict accurately the density profile deep
in the liquid region, where both soliton interactions and lattice
commensuration effects have a strong impact on the solution, Figure
\ref{tProfiles} gives an accurate representation of the commensurate
regions and their immediate vicinity.  Experimentally speaking, the
most interesting features of these profiles are the spatially
separated commensurate plateaux, which even at $t/V =0.02$ can cover
a significant fraction of the trap's volume, and hence should be
experimentally visible in the structure factor.

\section{Departures from Convexity}  \label{ConvexSect}

The PUH CGS are the classical ground states so long as the
potential is everywhere convex.  Since the on-site potential $U$
is tunable experimentally, it is interesting to ask what
happens to these states as $U$ is lowered away from convexity and
double occupancies begin to form. Of course, as $U$ is lowered still
further, triple and higher occupancies will also form.
However, as the barrier to triple occupancies is $3 U$, instabilities
towards triple occupancy at a given filling will set in at
approximately one third the value of $U$ for instabilities to
double occupancy. For $U$ in the range of relevant for 
interesting physics about the $1/2$-filled state which we will 
describe below, for example, triple
occupancies will not be favored at any filling fraction.
Hence we will not
analyze triply occupied states here---the doubly occupied r{\'e}gime has enough
challenges of its own.

As $U_0$ is deformed away from convexity, a series of
thresholds exists, at values of $U_0$ decreasing monotonically with the density of bosons.
At each (commensurate) density, we consider two thresholds.
The lower threshold is where double occupancies start to form spontaneously
in the CGS, and a new non-HUP classical ground state takes over
at this commensurate filling.
While our primary interest is in ground state transitions, much insight
is gained by also computing a second, upper threshold at a given filling. Past
this threshold,  $U=U_c^{(qSS)}$, a particle added to the state goes in
as a double occupancy instead of fractionalizing into $q$ solitons.  We will
see that these two thresholds allow us to understand many striking
features of the phase diagram.

\begin{figure}[htp]
\begin{center}
 \includegraphics[totalheight=10cm]{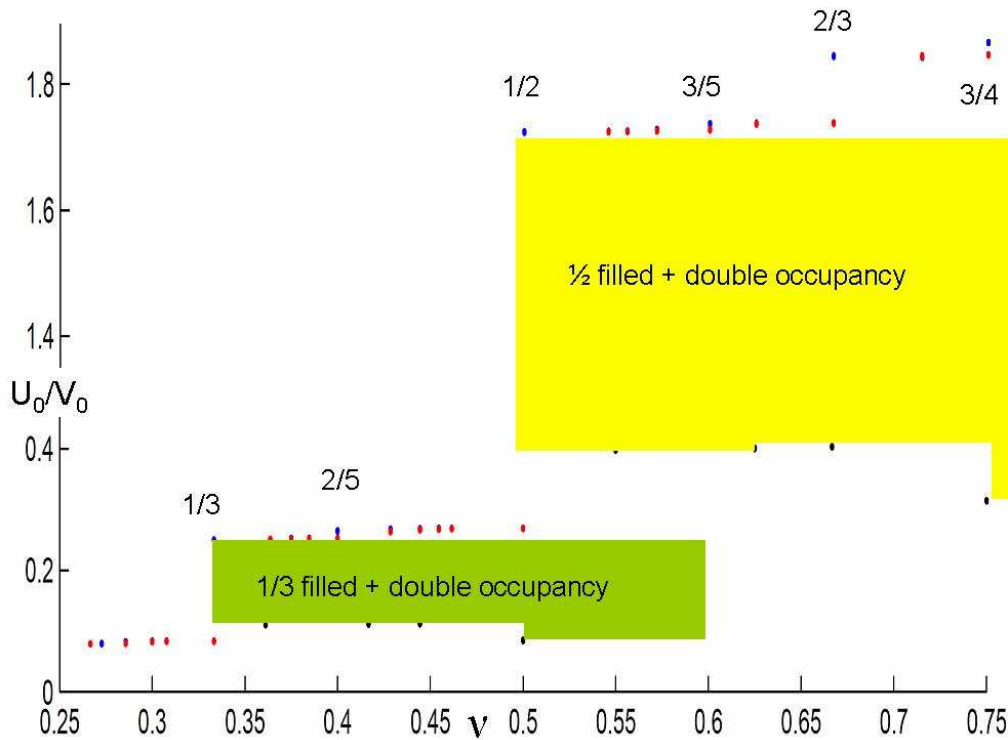}
\end{center}
 \caption[Thresholds of stability to double occupancy in the classical system]{\label{U0s} Numerically calculated values of $U_c^{(CGS)}$ (red) and
$U_{c}^{(qSS)}$ (blue), for a selection of filling fractions
$\nu$.  Quoted values of $U$ are measured relative to $V_0$.  A segment of the $y$-axis between $U_c^{(CGS)}(1/2)$ and
$U_c^{(qSS)}(1/2)$ has been removed for better resolution of the
rest of the phase diagram.  The green and yellow boxes, bordered
by black dots, indicate the approximate regions where the ground
states consist of double occupancies in the $1/3$ and $1/2$-filled
states respectively. Between the shaded regions the ground states
consist of double occupancies on other, higher-denominator states.
 }
 \end{figure}

 Fig.~\ref{U0s}
plots $U_c^{(qSS)}$ (blue) and $U_c^{(CGS)}$ (red) for states with
$q \leq 15$ in the vicinity of half-filling. The values of $U_c^{(qSS)}$
shown there are obtained by numerical minimization in the sector with one
added charge at the specified filling. The values of $U_c^{(CGS)}$ are
obtained by numerical minimization over configurations {\it at} the specified
filling that contain exactly
one double occupancy. The latter is the correct answer for all $\nu =p/q$
for which $p$ and $q$ are not both odd. In such cases the double occupancy
and its surrounding charge rearrangements give rise to even moments starting
with the equivalent of a quadrupole. Consequently, double occupancies repel at all distances
and enter via a continuous transition at the computed threshold. However,
when $p$ and $q$ {\it are} both odd, double occupancies have a dipole-like moment,
causing them to attract at long distances.
The transition in this case is first order, and the true $U_c^{(CGS)}$ lies
above our numerically determined value.
We ignore this gap due to first-order effects here, as we do not expect it to be
very large.  Indeed, a relatively straightforward calculation shows that the minimum in
the dipolar potential between two such defects (see Sect.  \ref{DipoleDef}) places
the two defects at least a distance $q$ apart; hence this energy gap decreases at least
as $\frac{1}{q^3}$.  This is a small perturbation in states of filling $p/q$ where $p$ and
$q$ are both large.

This section outlines the energetic arguments for the locations of these thresholds and
the nature of the new ground states, and
discusses the interesting features of Fig. \ref{U0s}.  We will end by exploring what can
be said about the phases deep in the non-convex regions -- which will lead to an interesting
new series of states which will form the subject of the last two sections of this paper.

\subsection{Effective convexity and thresholds for double-occupancy formation}

We begin by understanding where each CGS becomes unstable to
forming double occupancies.
Let $U_{c}^{(CGS)}(\nu)$ be the thresholds at which the PUH ground
states give way to ones with at least one double occupancy---these
are marked as the red points in Fig.~(\ref{U0s}). Observe that these
thresholds increase monotonically with $\nu$, $U_c^{(CGS)} ( \nu') >
U_c^{(CGS)} (\nu)$ for $\nu' > \nu$.

To understand this monotonicity, along with the approximate locations
of these thresholds, we considering the implications of
convexity at a given filling. Sufficient conditions for
convexity~\cite{Hubbard} are that, for all $x$,
 \ba \label{Convex}
\frac{1}{2} (V(0)+ V(2 x) )  \geq V(x)\ \ \ {\rm or} \ \ \
 U \geq\frac{15}{8} V_0 \left (\frac{1}{x^3} \right )
 \ea
For convexity to hold everywhere, (\ref{Convex}) must hold for
$x=1$; below this threshold double occupancies may occur. However,
when perturbing about a given convex solution at fixed $\mu$,
solutions will be stable approximately until $U$ violates
(\ref{Convex}) {\it for $x =r_{m}$}, the minimal inter-particle
distance. This implies that states with lower
filling fractions are more stable against double occupancies, as
the potential gain in lattice energy from doubly occupying a site
is smaller. This is in contrast to the stability of the
commensurate states as $t$ increases, where the denominator of the
filling fraction determines stability.

For example, consider the states $\nu <1/2$.
In reality, these states contain no pairs of particles
separated by $x=1$, so the convex solutions should be stable until
$U_0$ violates (\ref{Convex}) {\it for $x =2$} -- i.e. for $U_0 \geq
.235$.  More generally, we expect the Hubbard solutions to remain the ground
states so long as $U_0$ satisfies Eq. (\ref{Convex})  for $x \geq r_m$, the minimum
inter-particle spacing.

\subsubsection{Arguments for the approximate locations of the transitions}

For fillings $\nu = 1/q$ we can understand this more
rigorously by considering the energetics of nearest
neighbors. (The potential $1/r^3$ falls
off quickly enough that nearest-neighbor interactions
dominate the energetics in this case).
Since all particles in the initial CGS
have the same separation, forming a double occupancy in the CGS
results in replacing
\be \label{DOCGS}
 (q+1)V(q) \rightarrow (q+2) V(q+1) + U_0 \ \ \ .
\ee
Removing a
particle entirely changes the potential energy by $(q+1) V(q) - q V(q+1)$.  The double
occupancy, however, adds 2 NN distances of $q+1$ to the lattice: one
of these was originally a NN separation of $q$; the other was a NNN
separation of $2q$ -- which we omit here as it is a higher-order term.

Equation (\ref{DOCGS}) shows that the scale at which double
occupancies become energetically favorable in the CGS is
determined by $q$: if ( \ref{Convex}) holds for $x=q$, then double
occupancies will not form. Indeed, writing 
\ba q = \frac{q}{q+1}
(q+1) +(1- \frac{q}{q+1}) 0 \ea 
we see that for $V$ convex, 
 \ba
V(q) &=& V(\frac{q}{q+1} (q+1) + 0) \leq \frac{q}{q+1}V(q+1) +(1-
\frac{q}{q+1}) V(0) \n 
\mbox{or } (q+1) V(q) & \leq & q V(q+1) + U_0 
\ea 
Hence if
$V$ is convex {\em at the length scale set by $q$}, double
occupancies cannot be present in the classical ground state.

Similarly, we can gain an intuitive grasp for the energetics of $U_c^{(qSS)}$
by considering only nearest-neighbor interactions, and repeating the
above analysis.
If a soliton is formed in the $qSS$, the $q$ solitons  have replaced
\be \label{DOqSS}
(q-1)V(q) \rightarrow q V(q-1)
\ee
whereas inserting a particle as a double occupancy induces $2$
nearest neighbor distances of $q$.  The net NN difference in lattice
energy between adding a particle as solitons and adding a particle as
a DO is thus
\be
 \delta E  = (q+1) V(q)+U_0 - q V(q-1)
\ee
In this case, DO's will not form if Eq. (\ref{Convex}) is satisfied
for $x = q-1$.

Thus, we expect that the CGS and qSS will become unstable to
double occupancies at approximately \ba U < &U_c(CGS) \approx&
\frac{15 V_0}{8} \frac{1}{q^3} \n U< & U_c(qSS) \approx& \frac{15
V_0}{8} \frac{1}{(q-1)^3} \ea In short, states of higher density
will become unstable to double occupancies at larger values of
$U$, and at a given filling there is a finite gap between
$U_c^{(qSS)}$ and $U_c^{(CGS)}$.

What about more general filling fractions?  Repeating the above
analysis, we see that for fillings $p/q$, $p>1$, nearest neighbor
distances are $r, r+1$.  Double occupancies in the CGS replace
$(r+1)V(r)$ by $(r+2) V(r+1)$.  Likewise the energy cost of a
soliton is $(r+1)V(r)-r V(r+1)$, while inserting an extra particle
as a double occupancy adds $2 V(r+1)$.\footnote{Since $V(r)>
V(r+1)$ in the case of interest, we would expect this to be a
lower bound. In fact the double occupancy always polarizes the qSS
such that its two NN distances are exactly $r+1$.}  Hence the
nearest neighbor energy cost of forming a double occupancy in the
CGS and qSS is: 
\ba \label{pnot1} 
CGS & \delta E & =
(r+2)V(r+1)-(r+1)V(r) +U_0 \n qSS & \delta E & = (r+2)V(r+1) -
(r+1)V(r) +U_0 \ea 
Hence in general, we expect the approximate
value of $U_0$ at which these transitions occur to be set by the
smallest nearest-neighbor distance in the CGS.  However, the scale
of the difference between $U_c^{(qSS)}$ and $U_c^{(CGS)}$ is
determined by further neighbor interactions -- in fact, not until
$p^{th}$ neighbors are included in the energetic calculations is the
set of allowed separations different in the qSS and CGS.  Hence
this splitting will be much smaller in this case than for the
$1/q$ -filled state with $q=r$ in Eq. (\ref{pnot1}).

We now begin to understand the basic features of Figure \ref{U0s}.  To
a first approximation (obtained by considering only NN distances), we
have:
\ba
U_c^{qSS}(\frac{1}{q}) = U_c^{CGS}(\frac{1}{(q-1)})
\ea
and, for all $\nu$ with $\frac{1}{q} < \nu < \frac{1}{q-1}$,
\ba
 U_c^{CGS}(\nu) = U_c^{qSS} (\nu) = U_c^{qSS}(\frac{1}{q})
\ea
That is, the value of $U_0$ at which double occupancies first appear
is approximately decreasing with the filling.  There is a large gap
between its values for CGS and qSS states at filling fractions
$\frac{1}{q}$; the jump for fractions $\frac{p}{q}, p>1$ is much smaller.

\subsubsection{Form of the doubly occupied ground states and computing the exact loci of
transitions}  \label{UStates}

To calculate the exact loci of the transitions requires finding the exact occupation pattern
in the doubly occupied states, and computing its energy relative to that of states without double
occupancy.  The second part of this task is easy to do numerically, if not analytically; here we
discuss the energetically optimal configurations with double occupancy in both qSS and CGS states.

{\bf qSS states:}

We begin with the somewhat simpler qSS states.  To find the optimal configuration, we
first search for the optimal position in the repeating CGS pattern at which to add a double
occupancy.  Intuitively, this will be at the occupied site with the lowest local charge density.  Second, we ask whether adding a charge at this site creates further distortions
of the CGS state.  Though we will not prove that this is the globally optimal
configuration, the resulting configuration will give the least distortions of the
CGS configuration for a given local charge density, and hence should be the ground
state.

First, how do we identify the locus of lowest local charge density?  It is useful to consider a few simple examples:
\begin{itemize}
\item For states of the form $1/q$, there is only one site per unit cell -- hence no freedom in
where to place the double occupancy.  Using the notation of
Section \ref{HubbardSect}, we see that states of filling $2/q$
have occupancy patterns $...\ r \  r+1 \ r \ r+1 ... $, and again
both sites in the unit cell have the same local charge density.
\item At filling $3/q$, the occupancy pattern is either $ ... \ r \ r \  r+1 \ r \ r \ r+1 ... $, or $ ... \ r \ r+1 \  r+1 \ r \ r +1 \ r+1 ... $.  In the former case, the DO cannot sit between
the two NN distances of $r$; in the latter, it must sit between the two NN distances of $r+1$.
\item In the state $5/ 13 =  ...2 \ 3  \ 2 \ 3 \ 3 ... $, the DO must sit between the two NN
distances of $3$ -- which we will henceforth denote as the $5^{th}$ site in the unit cell, using
the convention that we count from left of the first interval shown.
\item In the state $5/ 12 =  ...2 \ 3  \ 2 \ 2 \ 3 ... $,
the DO must sit on the first or second
site in the unit cell (these being equivalent, up to reflection of
the entire pattern).  In this case, considering only nearest
neighbors would suggest that it could sit anywhere but the
$4^{th}$ site.  However, occupying sites $3$ and $5$ leads to
second neighbor distances of $4, 5$, instead of $5,5$ for sites
$1$ and $2$, making these energetically preferable.
\end{itemize}

This suggests the following simple algorithm: begin by considering
the sum of the two nearest-neighbor distances $s_1(i) =r_{i,l}+
r_{i,r}$ to the left and right of each site in the pattern.  If
one of the $s_1(i)$ is less than the rest, place the DO here.  If
not, generate the set of sums of right and left second neighbor
distances: $s_2(i) = r_{i, l} +r_{i-1, l} + r_{i, r}+ r_{i+1, r}$.
If this set has a unique largest element at site $i$, place the DO
there.  If not, proceed to the set of sums of right and left third
neighbor distances, and so on.  One can proceed in this way up to
$p-1^{st}$ neighbor distances; sites for which $s_k(i) =
s_k(j)_{k=1}^{p-1}$ have identical local charge density -- as in
the example above at $\nu = 5/12$-- and hence are related by
reflections and translations of the unit cell.

Having found the optimal site at which to add the DO, we must then
ask whether inserting charge here results in a further
re-arrangement of the background charge of the lattice.
Basically, dipolar `charges' on nearby sites are repelled by the
extra charge at the DO.  Charge will be forced away from the
doubly occupied site when the potential energy gain in doing so is
greater than the cost of compressing the state by the
corresponding amount in the remainder of the lattice.  When such
compression is favorable, it will result in particle-like solitons
forming near the DO and being repelled to infinite distance.  In
practice, this means that displacing one charge, say at the right
of the DO, one lattice site outwards will push all charges to its
right one site outwards as well.  This creates a single soliton at
$\infty$, as we have lengthened one $p^{th}$ neighbor distance
from $q$ to $q+1$.

It is straightforward to calculate the relevant energies.  The lattice energy associated
with a single soliton given by Eq. (\ref{Bak}):
$$\Delta E_1 = 1/q \sum_{p\neq 0 (\mbox{mod }q) }\left[ (r_p+1) V(r_p) - r_p V(r_{p}+1)
\right] +\sum_n nq V(nq-1) -(nq-1)V(nq) + ... \ \ \ .
$$
  We must compare this to the energy
gained by creating an extra hole between the DO, and site $i$, a
distance $d$ to its right (say). This energy has two
contributions: one infinite sum for the change in interaction of
site $i$ and all particles to its right with the double occupancy,
and one double infinite sum for the change in interaction of site
$i$ and all particles to its right with all occupied sites to the
left of the DO.  This gives: \ba   \label{Dshift_2} \Delta E_2
&=&\sum_{k=i}^{\infty} \frac{1}{ (x_k - x_i +d+1)^3} - \frac{1}{
(x_k - x_i +d)^3} \n &&+ \sum_{j=-\infty}^{i-1}
\sum_{k=i}^{\infty} \frac{1}{ (x_k + x_j - x_i +d+1)^3} -
\frac{2}{ (x_k+x_j - x_i +d)^3} \ea where $x_k$ here denotes the
position of the $k^{th}$ particle on the chain, relative to an
arbitrary origin.  The sum over $j$ here includes the doubly
occupied site; the first line of Eq. (\ref{Dshift_2}) counts only
the extra interaction due to the second particle occupying this
site.

The important point here is that the sum of the second line of Eq.
(\ref{Dshift_2}) and the soliton energy is positive-definite by
convexity of $1/r^3$.  Further, their sum depends only on which
particle in the unit cell lies at site $i$, rather than the value
of $d$ directly.  The gain in energy from moving away from the DO
(first line of Eq. (\ref{Dshift_2}), conversely, falls of rapidly
with $d$. Hence if $d$ is sufficiently large, the repulsion of the
DO will be too weak to push the charge outwards.  Further, once we
have found a site for which $\Delta E_1 + \Delta E_2 >0$, no
charge further from the DO than this site will be displaced.  It
is thus straightforward to compute $\Delta E_1 + \Delta E_2$ for
the sites close to the DO to determine which charges will be
pushed outwards.  The algorithm begins with sites closest to the
DO; if $\Delta E_1 + \Delta E_2<0$, the charge in question and all
charges further from the DO are pushed outwards by one site,
creating a single soliton infinitely far away.  One then moves
outwards to the next closest charges, and repeats the process
until a site is reached for which $\Delta E_1 + \Delta E_2>0$.

{\bf CGS States}

Forming double occupancies in the CGS is qualitatively different, since we think of re-arranging
the existing dipoles, rather than adding a new one.  In this case it is simplest to think
of forming the DO by {\em pinching} charge inwards towards an unoccupied site on which the DO
will form.

The pinching operation involves picking a site $x_d$ on which the
DO will be formed, and moving some number of particles (initially
at positions $\{R_i \}$) to the left of $x_d$ rightwards, and some
number of particles to the right of $x_d$ (initially at positions
$\{L_i \}$) leftwards.

For example, the $1/3$ state introduces $\{ L_i \} =
\{1\}$, and $ \{R_i \} = \{2, 5\}$, relative to $x_d$.  A DO can be formed in the $1/3$-filled
state as follows:
\ba
\frac{1}{3} &: &1 \ 0 \ 0 \ 1 \ \mathbf{0} \ 0  \ 1 \ 0 \ 0 \ 1 \ 0 \ 0 \ 1 \n
\mbox{ Pinch } L_1, R_1 &: & \ 1 \ 0 \ 0 \ 0 \ \mathbf{2} \ 0 \ 0 \ 0 \ 0 \ 1 \ 0 \ 0 \ 1 \n
\mbox{ Pinch }  R_2 &: & 1 \ 0 \ 0 \ 0 \ \mathbf{2} \ 0 \ 0 \ 0 \ \mathbf{1} \ 0 \ 0 \ 0 \ 1
\ea
Similarly, for the $2/5$ state,  $\{ L_i \} =
\{1, 4\}$, and $ \{R_i \} = \{1, 4\}$
\ba
\frac{2}{5} &: &1 \ 0 \ 1 \ 0 \ 0 \ \mathbf{1} \ \mathbf{0} \ \mathbf{1} \ 0  \ 0 \ 1 \ 0 \ 1 \ 0 \ 0 \ 1 \n
\mbox{ Pinch } L_1, R_1 &: & 1 \ 0 \ \mathbf{1} \ 0 \ 0 \ 0 \  \mathbf{2}  \ 0 \ 0 \ 0 \ \mathbf{1} \ 0 \ 1 \ 0 \ 0 \ 1 \n
\mbox{ Pinch }  R_2 &: & 1 \ 0 \ 0 \ 1 \ 0 \ 0 \ \mathbf{2} \ 0 \ 0 \ 1 \ 0 \ 0 \ 1 \ 0 \ 0 \ 1
\ea

Hence again, we must first find the optimal site onto which to pinch the charge, and then
ask whether other charges will move after the DO has formed.  The lowest energy final charge
configurations are as symmetric as possible: if the local charge density is lower on one
side of the DO, this inhomogeneous distribution will exert a force on the excess charge of the doubly occupied site.  Thus we find the optimal site by searching for holes which sit at symmetry
centers of the configuration.

This approach reveals an important even-odd type effect in forming
DO in the CGS.  If $q-p$ is odd, every hole in the unit cell
has an equal number of holes in the pattern to its right and
to its left, and we will show that it is possible to choose
a hole about which occupied sites are distributed symmetrically.
Pinching in these states will not alter this symmetry, and hence
two DO's repel.  If $q-p$ is even, however, such a site
does not exist and the final charge distribution is not symmetric.
This leads to a dipole-like moment for each double occupancy,
causing these to attract at long distances.

As a simple example of how this works, consider the states at $2/5$ and 
$3/5$ filling.  The occupancy patterns are:
\ba
\nu &= 2/5 \ :  \ \ \ & ... 1 0 1 0 0 1 0 1 0 0 ... \n
\nu &= 3/5 \ : \ \ \ & ... 1 0 1 1 0 1 0 1 1 0 ... \n
\ea
The important difference between the two is that at $\nu = 2/5$, 
there is an odd number of holes in the unit cell -- and hence 
somewhere in the unit cell  
there is a pair of occupied sites separated by an odd number of holes.  The distribution of 
charge required by convexity is such that the hole at the center of this interval 
is a center of symmetry of the charge distribution.  The double occupancy 
can then be formed by pinching particles onto this center of symmetry.  The ensuing 
charge distribution is therefore symmetric.  
At $ \nu = 3/5$, the center of symmetry of the pattern occurs on an occupied site, 
and hence there is no way to pinch the charge completely symmetrically.  The result is 
a configuration which has a dipole-like moment due to this charge asymmetry.

Indeed, in general HUP solutions will have a center of symmetry, since this is the 
most homogeneous way to distribute charge.  For general $p/q$ with $p-q$ odd, the unit cell 
contains an odd number of holes, and this center of symmetry consequently lies on an 
unoccupied site.  For $p-q$ even, conversely, it lies on an occupied site.  Thus 
the HUP solutions also result in the even-odd effect observed in the phase 
transitions to doubly occupied states.

This pinching construction gives a simple algorithm for constructing DO
in the CGS: first we find a hole which sits at a center of symmetry, and
pinch the two nearest bosons onto this site.  This can only create
attractive interactions from the remaining charges towards the DO.  We
then iterate through the neighboring particles and compute the energy
of moving each one inwards.  (Here one must bear in mind that for less
dense CGS states, it may be energetically favorable for particles to move inwards by
several lattice sites).  If this energy is negative, we move the particles inwards
and proceed to the next nearest occupied sites.  If the energy is positive, no
further re-distribution of charge will occur.

Hence for both CGS and qSS states, a simple algorithm exists to find the optimal
distribution of occupied sites, given that a double occupancy will form.  This
allows us to compute the exact difference in potential energy between
configurations with and without a single double occupancy -- and hence
calculate the threshold values of $U_0$ at which double occupancies
begin to form.  The result is the classical phase portrait Fig. \ref{U0s}.

\subsection{Arguments for monotonicity of $U_c$ with filling fraction}

 From
our above considerations, at fixed filling, the CGS is always more
stable against forming double occupancies than the qSS.  Indeed, a
simple estimate suggests that the gap separating these
instabilities is approximately the difference in energies for
adding and removing a particle to the state-- which is equal to
the interval in $\mu$ over which the state is stable at infinite
$U$. We now use this to understand an intriguing property of Fig.
\ref{U0s} -- namely, the monotonicity of $U_c^{(CGS)}$ with $\nu$.

The monotonicity can be understood as a simple consequence of the
fact that $U_c^{(qSS)}(\nu) > U_c^{(CGS)}(\nu)$.
Consider a filling $\nu' > \nu$. As $\nu'$ can be constructed
by adding charge to $\nu$, we conclude that $U_c^{(CGS)}(\nu') >
U_c^{(qSS)}(\nu)$; the latter corresponds to the threshold at
which adding an extra filling $\nu' - \nu$ in the form of the
solitons of $\nu$ loses out to adding it in the form of double
occupancies on top of the PUH state at $\nu$. As filling factors
only slightly greater than $\nu$ involve a dilute addition of
charges, we further conclude that $\lim_{\nu' \rightarrow \nu^+}
U_c^{(CGS)}(\nu') = U_c^{(qSS)}(\nu)$.

A somewhat more involved
argument shows that $\lim_{\nu' \rightarrow \nu^-}
U_c^{(CGS)}(\nu') = U_c^{(CGS)}(\nu)~$.  We have:
\be
\lim_{\nu' \rightarrow \nu^-} U_c^{(qSS)}(\nu')
= U_c^{(CGS)}(\nu)
\ee
 since the state at filling $\nu-\delta$ is the
state at filling $\nu$ with a density $~\delta$ of hole-like
solitons.  As this density decreases a single particle added as a
double occupancy will induce a charge configuration increasingly
similar to that of forming a double occupancy in the $p/q$-filled
state (which can be thought of as forming q hole-like solitons by
removing a single particle, then re-inserting this particle as a
double occupancy and letting the charge settle into its optimal
distribution.)  Further, the gap between $U_c^{(qSS)}(\nu')$ and
$U_C^{(CGS)} (\nu')$ vanishes as the denominator of the state
$\nu'$ goes to infinity, implying continuity of $U_C^{(CGS)}$ from
the left.

In short, our considerations so far imply intricate behavior for the location of
the initial ground state instability, namely that $U_c^{(CGS)}(\nu)$ is
a monotone increasing function of $\nu$ on the set of rationals with a
discontinuity at each rational value of $\nu$:
$$
\lim_{\nu' \rightarrow \nu^+} U_c^{(CGS)}(\nu') > U_c^{(CGS)}(\nu)
=\lim_{\nu' \rightarrow \nu^-} U_c^{(CGS)}(\nu').
$$
The first relation is a strict inequality, as the scale is set by the
 gap between $U_c^{(CGS)}(\nu)$
and $U_c^{(qSS)} (\nu)$, which is finite.  The second is equality,
because the gap is set by $\lim_{\delta \rightarrow 0} U_c^{(qSS)}
(\nu+\delta) - U_c^{(CGS)}(\nu+\delta)$, which is 0.

\subsection{Structure of the doubly-occupied r{\'e}gime}
\label{DOStrSect}

The previous sections have outlined some rather stringent
constraints on the thresholds $U_c^{(qSS)}$ and $U_c^{(CGS)}$,
which give significant insights into the nature of the
phase portrait of the classical system as $U_0$ is decreased:
monotonicity locates the transitions of higher denominator states
relative to those of lower denominator states.  Thus
we can understand the coarse features of this instability
by considering
first the most stable states (of denominator $q \leq q_0$, for some $q_0$ small
enough to allow the exact thresholds to be computed), and deducing
the expected behavior at fillings close to these.

We now turn to the question of what can be said about the phases
with $U < U_c$.
At any given filling, tracking the evolution of the ground state
with decreasing $U_0$ after double occupancies have been introduced is
a problem of considerable complexity.  Here we use the ideas
developed thus far to identify a family of regions in the $(\nu,U_0)$
plane where simpler descriptions emerge---these are indicated, in
two simple cases, by the shaded regions on the figure. The basic
idea is that once the $\nu = p/q$ qSS becomes unstable to double
occupancy, any particles added to the $\nu= p/q$ state will be added
as double occupancies, since these repel less strongly than
solitons (see below).  Hence at first sight we expect, in the region
$U_c^{(qSS)}(p/q) > U > U_c^{(CGS)}(p/q)$,  states of filling $\nu >
p/q$ to consist of double occupancies in the $\nu = p/q$ state.  At
rational fillings the double occupancies will arrange themselves in
a crystal thus generating a commensuration distinct from that of the
underlying $p/q$ state---we will refer to these as
doubly-commensurate states.

A detailed discussion of why, for $U_0 < U_c^{(qSS)}(\nu)$,
increasing the filling fraction can only form new double
occupancies, and never
 new solitons, is presented in Sect.  \ref{DOAdd_App}.  It is useful,
however, to understand the simple physical reason underlying this: adding particles
as solitons results in a charge distribution that is maximally spread out in space;
inserting them as double occupancies produces a maximally localized charge distribution.
If the change in density is infinitesimal, the solitons or double occupancies will
be infinitely far apart, and thus by definition if $U_0 < U_c^{(qSS)}(\nu)$ these
extra particles will enter the ground state as double occupancies.  If the change in
density is finite, we must add some number of particles per unit length on the lattice.
Since the potential $1/r^3$ falls of rapidly in space, the repulsion between two added
charges in a given distance is smallest when the extra charges are
as localized as possible --that is, when
both form double occupancies.  Hence as the density increases, so does the energetic
payoff of forming DO, rather than solitons.

Our discussion so far suggests that for all $\delta \nu >0$, states at filling
$\nu + \delta \nu$ with $U_c^{(CGS)}(\nu) < U_0 < U_c^{(qSS)} (\nu)$ consist of
the doubly commensurate states described above.  However, as the density of added charge
increases, the parent state itself becomes less
stable to forming extra double occupancies.  This can lead to
transitions in which the structure of the CGS collapses to a crystal
of double occupancies over a background of significantly smaller
filling.

We have carried out a simple analysis of the location of
this instability for the $1/2$ and $1/3$ plus double occupancy
regions in Fig.~(\ref{U0s}) at selected fillings and these are
marked by the black dots in the figure.  We do not
know of a general algorithm to compute this threshold at arbitrary
filling, but the principal is easy to illustrate for the $1/2$-filled
state.  Consider the following two states at filling $2/3 = 1/2 + 1/6$:
\ba
... \ 2 \ 0 \ 1 \ 0 \ 1 \ 0 \ 2 \ 0 \ 1 \ 0 \ 1 \ 0 \ ...  \n
... \ 2 \ 0 \ 0 \ 2 \ 0 \ 0 \ 2 \ 0 \ 0 \ 2 \ 0 \ 0 \ ...
\ea
The first state will be stable for $U_0 < U_c^{(qSS)} (1/2)$, but sufficiently large.
However, the energy gained by pinching this state to form the second state is
greater than the energy gained by simply forming a DO in the $1/2$-filled state,
as the two particles that are pinched are moving away from sites with charge $2$,
rather than charge $1$.  The black dots in the figure are calculated by calculating
the difference in energies of such simple configurations.

It is interesting to notice that the black dots do not behave monotonically with filling
fraction.  To understand why, consider the following two states at filling $3/4$:
\ba
... \ 2 \ 0 \ 1 \ 0 \ 2 \ 0 \ 1 \ 0 \ 2 \ 0 \ 1 \ 0 \ ...  \n
... \ 2 \ 0 \ 0 \ 2 \ 0 \ 0 \ 2 \ 0 \ 2 \ 0 \ 0 \ 2 \ 0 \ 0 \ 2 \ 0 \ ...
\ea
Because of the greater density in the first state, the best we can do by
pinching is ensure that $2/3$ of the double occupancies sit at least $3$
lattice sites apart.
Hence once at least
half the sites are doubly occupied, the potential gain due to
forming extra double occupancies is actually less than for
smaller fillings.

In summary, Fig. \ref{U0s} gives an accurate picture of the locations
of the initial phase transition due to deforming $U_0$ away from
convexity for arbitrary fillings.  We have constructed the general
form of the resulting states with double occupancy.  As $U_0$ is
decreased even further, our analysis suggests that there are
further transitions to states with a higher density of double
occupancies, whose structure we have not systematically understood.
While we have found sizeable regions which can be
described as simple descendants of the $1/2$ and $1/3$ states, we
are not at present able to estimate the sizes of analogous regions
for higher denominator fractions. Of course, to have a full solution
of this would be equivalent to tracking the evolution of each $\nu$
as $U$ is decreased from $U_c^{(CGS)}$.

\section{Interesting phenomena in the non-convex r{\'e}gime }  \label{SSSect}

In Section \ref{ConvexSect}, we have expended considerable effort
understanding the various phase boundaries in the classical
system as $U_0$ is decreased away from the convex limit.
We now focus on a particularly interesting region of this
phase diagram -- namely, that in which the ground
states are derived by adding double occupancies to a parent CGS configuration.
We will first discuss the classical limit of these phases,
showing that these contain a re-scaled version of the devil's staircase.
Focusing on the devil's staircase near $1/2$ filling, we then consider
the effect of adding hopping to the mix, and find a phase diagram with
super-solid like regions in which a Luttinger liquid of double occupancies
co-exists with a commensurate $1/2$-filled background.

\subsection{ A new staircase}

The discussion of Sect. \ref{ConvexSect} has led us to
the doubly-commensurate states in the $(\nu,U_0)$ phase diagram: we
remind the reader that such states are constructed by periodically
doubly occupying some fraction of the sites in a CGS. We now discuss
the ground state configurations of these double occupancies, and show
that in at least some cases these doubly commensurate states can
form a devil's staircase of their own.

To explore such states, we will work in the parameter range where
increasing density effectively adds double occupancies to a parent
CGS state, without forming DO in the parent state itself.  We have
argued in Sect. \ref{DOStrSect} that at least for the $1/2$ and $1/3$
filled states, this description is apt over sizeable regions
of the phase diagram.  Though we have not calculated the
lower thresholds for higher denominator states, we expect that
a similar description holds for arbitrary parent
fillings $\nu$ and $U_c^{(CGS)} < U_0 < U_c^{(qSs)}$,
 at least over a modest
range of densities.

First consider states constructed from double occupancies in the
$1/2$-filled state, which exist in the region shaded in yellow in
Fig.~(\ref{U0s}). The energetics of such states can be divided into
a) the constant interaction of the parent $1/2$-filled PUH
configuration with itself, b) the constant interaction of the added
charges, irrespective of their location, with the parent $1/2$
filled configuration and c) the interaction of the added charges
with themselves. This last part involves an interaction between the
added charges which is convex again and thus leads to PUH
configurations sitting on a lattice with a doubled lattice constant.
The energy cost of adding a single double occupancy is $U+V_d$,
Hence the doubly-occupied sites comprise a Devil's staircase with
$\mu \rightarrow \mu +U + V_d$, and the widths of all intervals
decreased by a factor of 8. Here $V_d
=\frac{1}{8}\sum_{n=1}^{\infty} \frac{1}{n^3}$ is the interaction
energy of each double occupancy with the underlying $1/2$-filled
state.  At fixed $U$, this staircase is complete over the range of
fillings for which increasing the particle density infinitesimally
does not induce `excess' double occupancies to form in the
half-filled background lattice.  In the case of the $1/2$-filled
state, for $U$ sufficiently close to the upper cutoff this gives a
complete staircase on $1/2 \leq \nu \leq 1$.

Similar structures
exist for all $1/q$-filled states in the appropriate range of $U$.
As mentioned before, we do not, at present, understand the situation
for doubly commensurate descendants of general rational fillings.

\subsection{ Supersolids}
Thus far our considerations away from the convex limit have been purely
classical. But we can equally consider states obtained from these modified
classical states upon the introduction of hopping. Specifically, let
us consider the fate of the doubly commensurate descendants of the PUH
$1/q$ states considered above.

In a manner entirely analogous to the problem with which we began
this paper, the superlattice of added charges can melt via the
motion of {\it its} solitons as $t$ is increased resulting in a
phase transition between the doubly commensurate state and a
``super-solid'' like phase in which the background $1/q$-filled CGS
coexists with a Luttinger liquid. This is, in a sense the $d=1$
version of the supersolid in higher dimensions, but it is worth
noting that the $d=1$ version in our problem exhibits a more
divergent CDW susceptibility than superfluid susceptibility as $T
\rightarrow 0$.

\begin{figure}[htp]
\begin{center}
 \includegraphics[totalheight=10cm]{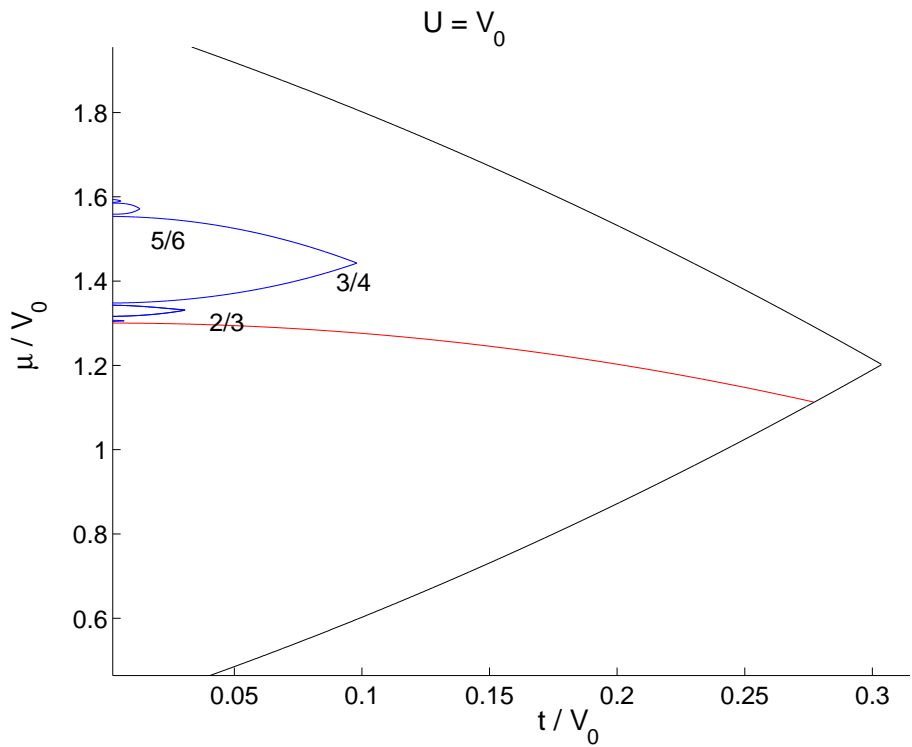}
 \end{center}
\caption[Super-solidity in the doubly occupied r{\'e}gime]{\label{SuperSolid} Phase portrait in the
vicinity of doubly commensurate states about $1/2$ filling at $U
=V_0$.  Again, $\mu, t,$ and $U$ in the figure are measured relative to $V_0$.
The black lines indicate the boundaries of the Mott lobe at
$U = 20$.  The red curve shows the chemical potential at which
it becomes energetically favorable to add particles to the
half-filled state as double occupancies, for $U=V_0$.  The blue
curves show some Mott lobes of the doubly-occupied staircase
region; the region between these and the black lines (which
delineate the region of stability of the $1/2$-filled state at
infinite $U$) is a super-solid state. }
 \end{figure}

To get a more quantitative account of these new phases, we may
repeat the strong coupling treatment above. Fig.~(\ref{SuperSolid})
shows the phase portrait at intermediate values of $U$ near
$\nu=1/2$. The black line traces the infinite-$U$ Mott lobe, over
which the background $1/2$-filled state is stable against forming
solitons. The red line shows the threshold at which it is
energetically favorable to add a single double occupancy to the
$1/2$ filled state.  The blue curves show the positions of the Mott
lobes for the doubly commensurate states. The presence of double
occupancies stabilizes the $1/2$-filled state against proliferation
of solitons, so that the background remains commensurate at least
within the infinite-$U$ $1/2$-filled Mott lobe, shown in black.
This $1/2$-filled super-solid phase has also been shown to exist in
an extended Bose-Hubbard model with second-neighbor
repulsion~\cite{ScalletarSuperSolid}; these numerical results are
consistent with the phase portrait shown here for small $t/V$.

\subsubsection{Perturbation theory in the super-solid r{\'e}gime }

The phase diagram in Figure \ref{SuperSolid} is obtained using
$4^{th}$ order perturbation theory in $t$. As the calculation of
the energy corrections is rather involved, we have included the
details in Sect.  \ref{SSPert_Sect}.  Here we will discuss the
form of these corrections, and the qualitative features of the
phase diagram.

The calculation is similar to that described in the convex case in Sect \ref{PertSect}, with
one key difference: all ground states must have holes at every second site.  Adding a charge
to a doubly commensurate crystal produces an extra double occupancy -- which in turn
produces solitons {\em in the doubly occupied sites}.  That is, as we do not allow 
triple occupancies to form, and it is energetically unfavorable to place extra 
particles on the unoccupied sites, we may {\em only} add particles by forming 
solitons in the doubly-occupied superstructure.  At filling $1/2 + p/2q$, for example,
the relevant ground state is given by the qSS of the $p/q$ filled state, with
all distances stretched by a factor of $2$.
Hence
 only even-order corrections to the energy are non-vanishing:
an odd number of hoppings produces a state with at least one particle on a site that is
vacant in the parent half-filled state.

As discussed in the Sect. \ref{Appendix}, we can therefore describe the
corrections in terms of an `effective' hopping Hamiltonian,
$H_1^{eff}$, which describes the hopping between occupied sites.
$H_1^{eff}$ contains both a `trivial' correction, in which a boson
hops off a given site and back on, and a `hopping' term in which a
boson hops from one occupied site to another. The second-order
correction to the energy in this case is: \be E^{(2)} = -
\langle\psi_0| H_1^{eff} |\psi_0\rangle \ee The trivial part of
$H_1$ gives approximately the same correction for commensurate and
solitonized states (except for small differences in the energy
denominators). The hopping term, on the other hand, has a
first-order matrix element with solitonized states, but not with
the CGS.  Thus the situation is analogous to that of qSS solitons
in the convex limit: solitonized states have a negative energy
correction at leading order in perturbation theory which their
commensurate brethren do not.  Thus as $t$ increases the double
occupancies undergo a commensurate-incommensurate transition, as
seen in Fig. \ref{SuperSolid}.

\section{ Concluding Remarks}

The infinite range of the dipolar interaction
does produce, as promised, intricate phase diagrams for the one
dimensional dipolar bosonic gas. Particularly striking are the singular
staircase functions that showed up in our analysis in three different
settings: in the $\nu(\mu)$ curve for the exactly dipolar classical
problem, in the function $U^{(CGS)}_{0c}$ which marks the instability
of the PUH states when the onsite $U$ is tuned down and in the
$\nu(\mu)$ curves in selected regions of the $(\nu, U)$ plane. The
other main set of results pertain to the presence of a large, indeed,
infinite number of transitions between Mott crystals and Luttinger
liquids or supersolids. The challenge of observing some of this physics
in cold atomic gases is not trivial---the major obstacles are getting
a reasonable simulacrum of a one dimensional gas of infinite extent. On
the positive side, the control parameters we study here are eminently
tunable.

\section{Supplementary Material} \label{Appendix}

\subsection{General Perturbation Theory} \label{PertApp}

Here we review the perturbation theory pertinent to treating the hopping terms in the general Hamiltonian, and in particular hopping in the super-solid state.

Let $H_0$ be the unperturbed Hamiltonian, with eigenstates
$|\psi_i^{(0)}\rangle$ of energy $E_i^{(0)}$.  The perturbation
$\delta H_1$ induces corrections to the wave function and
energies; the $i^{th}$ order corrections are denoted
$|\psi^{(i)}_j\rangle, E_j^{(i)}$.  When the lower indices are
omitted, we will be referring to corrections to the ground state
energy and wave-function (which is all we're interested in here.)

Perturbation theory can be summarized by the following equations:
\ba |\psi\rangle &=& |\psi^{(0)}\rangle + \delta|\psi^{(1)}\rangle
+ \delta^2 |\psi^{(2)}\rangle  + \delta^3 |\psi^{(3)} \rangle +
...\n E &=& E^{(0)} + \delta E^{(1)} + \delta^2 E^{(2)} + \delta^3
E^{(3)} + ... \n (H_0 + \delta H_{1} ) |\psi\rangle &=& E
|\psi\rangle \ea 
%
We stipulate that $\langle\psi_0^{(0)} |\psi^{(i)}\rangle =
\delta_{i0}$, 
%
and find the recursion relation for the energy is:
\be \label{Pcors} 
(E_i^{(0)} - E_0^{(0)})
\langle\psi_i|\psi^{(n)}\rangle = \left( \sum_{j=1}^{n-1}
E^{(n-j)} \langle\psi_i|\psi^{(j)}\rangle \right) -\langle\psi_i|
H_1 |\psi^{(n-1)}\rangle \ee
where $| \psi_i \rangle$ label excited states of the unperturbed Hamitlonian.

We can explicitly write out the first few terms of these series in terms of the unperturbed wave functions and energies.  We drop all superscripts, which are $0$.
\ba
E^{(1)} &=& \langle\psi_0 | H_1 | \psi_0 \rangle \n
|\psi^{(1)}\rangle &=&- \sum_i \frac{1}{ E_i - E_0} |\psi_i\rangle \langle\psi_{i} | H_1 | \psi_{0} \rangle \n
E^{(2)} &=& -\sum_i \frac{1}{ E_i - E_0 } |\langle\psi_0 | H_1 | \psi_i \rangle|^2 \n
|\psi^{(2)} \rangle&=&\sum_{j} \frac{1}{ (E_j - E_0)} |\psi_j\rangle\left ( \sum_i \frac{1}{E_i - E_0} \langle\psi_j| H_1 |\psi_i\rangle \langle\psi_{i} | H_1 | \psi_{0} \rangle \right. \n
&& \left.- \frac{1}{E_j - E_0} \langle\psi_0|H_1|\psi_0\rangle \langle\psi_j|H_1| \psi_0 \rangle \right )\n
E^{(3)} &=&\sum_j\left(- \frac{ \langle\psi_0|H_1|\psi_0\rangle}{ (E_j - E_0)^2 } |\langle\psi_0 | H_1 | \psi_j \rangle|^2  \right. \n
&& \left. +\frac{1}{(E_i-E_0)(E_j-E_0)} \langle\psi_0|H_1|\psi_j\rangle\langle\psi_j| H_1 |\psi_i\rangle \langle\psi_{i} | H_1 | \psi_{0} \rangle \right )
\ea
The matrix elements $\langle\psi_i| H_1|\psi_j\rangle$ are all $\pm t$ or $0$, except for certain hoppings in the vicinity of the soliton, which have the form $t(1+ \cos(kq))$.

We are interested in corrections higher than $3^{rd}$ order to investigate possible super-solid states, which consist of a background of filling $1/n$, and double occupancies at filling $p/(nq)$.  The non-vanishing energy corrections occur at multiples of $n$, as other orders cannot map the ground state back to itself.

Here we consider the simplest case, $n=2$.  Since the first order
correction to $E$ vanishes, we may write $\psi^{(2)}$ in the form
\be |\psi^{(2)} \rangle=\sum_{j} \frac{1}{ (E_j - E_0)}
|\psi_j\rangle \langle\psi_j| H_1^{eff} | \psi_{0} \rangle \ee
where $H_1^{eff} =  \sum_i \frac{1}{E_i - E_0} H_1 |\psi_i\rangle
\langle\psi_{i} | H_1$ is the effective hopping Hamiltonian. Here
$|\psi_i \rangle$ has one occupied odd site.   $|\psi_j\rangle$
differs from the ground state by $2$ hoppings.
 There are two such terms: $|\psi_{j_0} \rangle$ has occupation of even sites only, and $|\psi_{j_1} \rangle $
has two occupied odd sites.  It is useful to separate these two
terms explicitly: \be |\psi^{(2)} \rangle=\sum_{j_0} \frac{1}{
(E_{j_0} - E_0)} |\psi_{j_0}\rangle \langle\psi_{j_0}| H_1^{eff} |
\psi_{0} \rangle + \sum_{j_1} \frac{1}{ (E_{j_1} - E_0)}
|\psi_{j_1}\rangle \langle\psi_{j_1}| H_1^{eff} | \psi_{0} \rangle
\ee The first piece looks like the first-order correction to a
wave function on a re-scaled lattice, with an effective hopping
coefficient given by $H_1^{eff}$.  The second piece does not
contribute to the second-order energy correction.  Hence in this
notation, we also have \be E^{(2)} = - \langle\psi_0| H_1^{eff}
|\psi_0\rangle \ee This is the leading-order energy correction.
The third-order correction to the energy vanishes; the
wave-function correction is: \ba |\psi^{(3)} \rangle &=& \sum_i
\frac{1}{E_i-E_0} |\psi_i\rangle \left (\frac{1}{E_i -E_0}
\langle\psi_i| H_1| \psi_0\rangle
\langle\psi_0|H_1^{eff}|\psi_0\rangle \right.\n &&\left.-\sum_j
\frac{1}{E_j -E_0}\langle\psi_i| H_1 | \psi_j\rangle
\langle\psi_j| H_1^{eff} | \psi_{0} \rangle \right) \ea This gives
the fourth order energy correction: \ba E^{(4)} &=& \sum_i
\frac{1}{E_i-E_0} \langle\psi_0|H_1|\psi_i\rangle\langle\psi_i|H_1
\left (|\psi_0\rangle \langle\psi_0|H_1^{eff}|\psi_0\rangle
\frac{1}{E_i -E_0 } \right.\n && \left.-\sum_j |\psi_j\rangle
\langle\psi_j| H_1^{eff} | \psi_{0} \rangle \frac{1}{E_j-E_0}
\right) \n &=&   \langle \psi_0 |\tilde{H}^{eff}
|\psi_0\rangle\langle\psi_0|H_1^{eff}|\psi_0\rangle -\sum_{i_1}
\frac{1}{E_{i_1}-E_0}| \langle\psi_0| H^{eff}| \psi_{i_1}
\rangle|^2 \n && -\sum_{i_0} \frac{1}{E_{i_0}-E_0}| \langle\psi_0|
H^{eff}| \psi_{i_0} \rangle|^2 \ea where $\tilde{H}^{eff} = \sum_i
\frac{1}{(E_i -E_0)^2} H_1 |\psi_i\rangle\langle\psi_i| H_1$.  The
last term is the analogue, in our effective theory, of a
`second-order' correction.  The first two terms give an extra
correction due to the structure of the underlying lattice.

\subsubsection{Matrix Elements and formulae used for numerics in the super-solid
states}   \label{SSPert_Sect}

To evaluate the perturbative energy corrections, we must calculate the relevant matrix elements.  Here we outline this calculation order by order for the super-solid phases, in which the
results are somewhat complex.  The corresponding calculations for the convex r{\'e}gime are
a straightforward adaptation of the results of \cite{FreericksMonien}, and are not included
here.  To simplify the language, for the remainder of this section we will use qSS and CGS
to refer to states with and without solitons {\em in the doubly occupied sites}, respectively.

{\bf Second Order correction}

The second-order energy is given by
\be
E^{(2)} = - \langle\psi_0| H_1^{eff} |\psi_0\rangle
\ee
The matrix elements
\be
\langle\psi_0 | H_1 |\psi_i \rangle
\ee
are $\sqrt{2} t$ when $|\psi_i\rangle$ involves hopping from a doubly occupied site (as there are two identical particles to choose from), and $t$ otherwise.  (This is because $b_i | n_0\rangle = \sqrt{n_0} |n_0-1\rangle$.  A factor of $\sqrt{2}$ is also introduced by hopping onto an occupied site, since $b^\dag_i | n_0 -1 \rangle = \sqrt{n_0} |n_0\rangle$.  Hence hopping a particle off of a doubly occupied site, and then back on, incurs a factor of $2$).

Hence, for the CGS, \be E^{(2)} = -2 t^2 \left [ \sum_{i \neq
i_{DO} } \frac{1}{E_i - E_0} +  \sum_{ i_{DO} } \frac{2}{E_i -
E_0} \right ] \ee where $E_i$ corresponds to hopping the $i^{th}$
particle to the right.  (The overall factor of $2$ then accounts
for the left hoppings).  This can be written conveniently in the
form: \be E^{(2)} = -2  \sum_{i } t^{eff}_{i, R} \ee where
$t^{eff}_{i, R} = \frac{t}{E_i - E_0}$ if site $i-1$ is singly
occupied, and $ \frac{2}{E_i - E_0}$ if it is doubly occupied.
This accounts for the denominator when hopping a particle from
site $i-1$ one site to the right, to site $i$.  We may equally
well define $t^{eff}_{i, L}$, which is $\frac{1}{E_i - E_0}$ if
site $i+1$ is singly occupied, and $ \frac{2}{E_i - E_0}$ if it is
doubly occupied.

In the solitonized ground states, there is an additional contribution, in which the particle at one end of the soliton hops two sites to the right (or left) effectively hopping the soliton by $2q$ sites.  This extra contribution is equivalent to the first order energy gain of the solitonized state in the non-super-solid phase.  The $t^{eff}_i$ for this is the same as for this particle hopping out and returning to the same site.  Thus for the qSS
\be \label{EsolEx}
E^{(2)}_{qSS} = -2  \sum_{i } t^{eff}_{i, R} -2 t^{eff}_{i_s} \cos(2 k q)
\ee
where $t^{eff}_{i_s}$ contains the appropriate energy denominator for the hopping from site at the end of the soliton.  The factor of 2 in front of the sum accounts for left and right hoppings; in the second term, the particle can hop from the right end of the soliton ending at site $i_s$ to the left end of a (new) soliton beginning at $i_s+2$, or vice versa (
reversing left and right).  The difference between the CGS and qSS energy corrections is thus:
\be
\delta E^{(2)} = -2 \sum_{i=1 }^N \left[ t^{QSS}_{i, R}- t^{CGS}_{i, R} \right] -2 t^{eff}_{i_s} \cos(2 k q) - \frac{2 (p\pm1)}{p} \sum_{i=1}^p t^{CGS}_{i, R}
\ee
Here $N$ is the number of particles on the chain.  In practice the sum may be truncated at finite $N$ as the difference
between CGS and qSS contributions falls off rapidly far from a soliton.  In this term, however, we have assumed that there is one
particle in the CGS for each particle in the qSS; we must correct for this by subtracting the (average) energy of one particle in
the CGS, given in the last term.

{ \bf Fourth order correction}

The fourth order energy correction is: \ba \label{4thOrd} E^{(4)}
&=&   \langle \psi_0 |\tilde{H}^{eff}
|\psi_0\rangle\langle\psi_0|H_1^{eff}|\psi_0\rangle -\sum_{i_1}
\frac{1}{E_{i_1}-E_0}| \langle\psi_0| H^{eff}| \psi_{i_1}
\rangle|^2 \n &&-\sum_{i_0} \frac{1}{E_{i_0}-E_0}| \langle\psi_0|
H^{eff}| \psi_{i_0} \rangle|^2 \ea The first term in the sum is a
product of $E^{(2)}$, calculated above, and $\tilde{E}^{(2)}$,
given by \be \tilde{E}^{(2)} = -2 t^4 \left [ \sum_{i \neq i_{DO}
} \frac{1}{(E_i - E_0)^2} +  \sum_{ i_{DO} } \frac{2}{(E_i -
E_0)^2} \right ] \ee Here again, the solitonized ground state has
an extra contribution compared to the CGS, of the form
(\ref{EsolEx}).  Each term in this sum is uniquely labeled by
choosing two site indices, $i$ and $j$, to be the target sites of
intermediate hopping, and a direction of hopping onto and off of
each.

In the second term, $|\psi_i\rangle$ are states which are connected to $|\psi_0\rangle$ by hopping two particles $k$ and $j$ onto odd sites.  The choice of $k$ and $j$ uniquely fixes the intermediate state $|\psi_{i_1} \rangle$.  Note, however, that for each choice
of $k$ and $j$ there are two possible hopping sequences: $|\psi_{0} \rangle \rightarrow |\psi_{k} \rangle \rightarrow |\psi_{i} \rangle \rightarrow |\psi_{j} \rangle \rightarrow |\psi_{0} \rangle$, and $|\psi_{0} \rangle \rightarrow |\psi_{k} \rangle \rightarrow |\psi_{i} \rangle \rightarrow |\psi_{k} \rangle \rightarrow |\psi_{0} \rangle$, where $|\psi_k \rangle$ denotes
the state in which particle $k$ has hopped, but particle $j$ has not.  These two
possible paths differ only in the order in
which the hopping back to the ground state is performed.

Hence the second term of (\ref{4thOrd}) can be expressed as:
\ba
 -\sum_{i\neq j} n_i n_j \left (\frac{1}{ (E_{i+1, j+1} -E_0) (E_{j+1}-E_0)^2} + \frac{1}{(E_{i+1, j+1} -E_0) (E_{i+1} -E_0) (E_{j+1} -E_0)}  \right ) \n
- \sum_i n_i^2 \left(\frac{1}{ (E_{i+1} -E_0) (E_{j+1}-E_0)^2} + \delta_{n_i, 2} \frac{1}{(E_{i-1, i+1}-E_0)(E_{i+1}-E_0 )(E_{i-1}-E_0)}  \right ) \n
\ea
Here $i$ and $j$ are sites which are occupied by $n_i$ and $n_j$ bosons in the ground state.  $E_{i+1}$, $E_{j+1}$ are the energies
of the intermediate states produced by hopping a boson off of sites $i$ and $j$ (by hopping to the right; the term for hopping to the left appears with a $-$ sign).
$E_{i+1, j+1}$ represents the energy of the intermediate state with particles hopped from both sites $i$ and $j$.  Here again, in the case of solitonized states the particle at the
end of the soliton can hop either `out and back', or `out and out' -- the latter producing a translation of the soliton.

For both CGS and qSS states, the sum of the first two terms in Eq. (\ref{4thOrd}) is small.
In particular, the cross-terms between hoppings of the soliton itself and hoppings of the
background lattice approximately cancel.  This is as it should be -- if they did not, these
terms would give an energy splitting between qSS and CGS states which diverges in the
thermodynamic limit!

The final contribution resembles a second-order contribution with a re-scaled lattice constant and hopping terms which we will call $E^{(4)}_a$ and $E^{(4)}_b$.  The first of these has the form:
\be
E^{(4)}_a= -2 \sum_i \frac{ |t^{eff}_{i-2, R} |^2 }{E_i -E_0} \n
\ee
where $i$ labels the locus of the hopped particle in the excited state, and $i-2$ its locus in the ground state (prior to hopping).  The factor of 2 accounts for hoppings to the left and right.  The $t^{eff}_{i-2, R}$ is the effective hopping from site $i-2$ to site $i-1$ (which gives the same energy denominator as a leftward hop from site $i$ to site $i-1$).  As in the non-super-solid case, there is one special intermediate configuration $| \psi_i^*\rangle$ in the solitonized case which permits two distinct hoppings which return to the ground state, as the ground state in this case is a momentum eigenstate.  The intermediary configurations for the `forward' and `backward' hoppings are mirror images of each other in this case.  Hence for the solitonized ground state:
\be \label{E4Sol}
E^{(4)}_a= -2 \left [ \sum_{i\neq i_s} \frac{ |t^{eff}_{i-2, R} |^2 }{E_i -E_0} +   \frac{ |t^{eff}_{i^*} |^2}{ E_{i^*} - E_0} \right ]
\ee
We exclude hopping to the right from $i_s$ in the first sum, as it merely hops the soliton, and
has been included at second order.  The special hopping from site $i^*$ gets a matrix element of $4$, as it also occurs once in the first sum.

The final contribution represents the effect of hopping both particles off of a
doubly occupied site and then back, and has the form
\be
E^{(4)}_b = \sum_{i_{DO}} 4 \frac{ 2t^2 | t^{eff}_{i, R}|^2  }{E_{i, RR} -E_0} + 2 \frac{ 2t^2 }{E_{i, RL} -E_0} \left (  | t^{eff}_{i, R}|^2 +  |t^{eff}_{i,R} t^{eff}_{i,L}| \right )
\ee
where $E_i$ is the energy of the intermediate configuration with both particles removed from the original site.  Here one factor of 2 is, as usual, to account for both left and right hoppings.  In the first term, both particles are hopped to the same site; the factor of $4 = (\sqrt{2})^4$ arises because all hops move a particle either onto or off of a doubly occupied site.  In the second term, one particle is hopped to the right and one to the left; consequently there is only a single factor of 2 due to double occupancies.  However, the return hop may be executed in the same order as the outward hop ($|t^{eff}_{i, R}|^2$ term), or the opposite order ($|t^{eff}_{i,R} t^{eff}_{i,L}|$).  The form of this correction is the same for the CGS and for solitonized states.

\subsection{Bounds on the volume of Devil's staircase lost at small
$t$} \label{QScale}

Here we estimate the total volume of the liquid states in the phase diagram for small
but finite $t$.

First, let us review the situation at $t=0$, where the total
volume of liquid states is $0$. The range of stability of a Mott
lobe with denominator $q$ is given by \be \sum_{n} nq \left(
V(nq+1)+ V(nq-1) -2 V(nq) \right ) \ee To calculate the volume
occupied by all rationally filled states, we sum over q.  The
multiplicity of each $q$ is the number of rationals between $0$
and $1$ with denominator $q$, which is given by the Totient
function $\phi(q)$.  This gives the volume: \be \label{Summs}
\sum_q \phi(q) \sum_n  [ qn V(qn+1) + qn V(qn-1) - 2 qn V(qn) ]
\ee Switching the order of summation in the first term, and
letting $m = qn+1$, gives: \be \sum_m \left[ \sum_{q | m-1}
\phi(q) \right ]    (m-1) V(m) \ee The other terms in
(\ref{Summs}) can be treated similarly.  The expression in square
brackets is just $(m-1)$, so that the total is \be \sum_m [
(m-1)^2 +(m+1)^2 - 2 m^2] V(m) = 2 \sum_m V(m) \ee which is in
fact the threshold of stability (from below) of the integer-filled
state.

Now let us calculate the amount of this volume lost to superfluid
states at small $t$.  Since the form of the perturbative
corrections is difficult to deduce exactly in $q$, we work to
linear order in $t$.  Hence from each Mott lobe of denominator
$q$, a swath in chemical potential of length $2 q t$ has been lost
to superfluidity. Summing over $q$ gives: \be \Delta_1 = 2 t
\sum_{q\leq q_0} q \phi(q) \ee To avoid subtracting off an
infinite correction, we sum only over those values of $q$ for
which the Mott lobe has not, at this value of $t$, entirely
disappeared.  The threshold value of $q$ is given by: \ba 2 t q_0&
=& \sum_{n} \frac{nq}{(nq+1)^3}+ \frac{nq}{(nq-1)^3} -2
\frac{nq}{(nq)^3} \n &=& \frac{\pi^2}{3 q^3} \left[q( 2+ 3
\cot(\pi/q)^2) - 3\pi\cot(\pi/q)(1 + \cot(\pi/q)^2 ) \right] \ea
At $t =0$, $q_0$ is infinite; hence for small $t$ $q_0$ is very
large, and we may expand the right-hand side to leading order in
$1/q_0$, giving: \be t = \frac{\pi^4}{15 q_0^5} \ee

To complete the calculation, we must also estimate the volume of
the lost Mott lobes: \be \Delta_2 = \sum_{q>q_0} \sum_n \phi(q)
\left(\frac{nq}{(nq+1)^3}+ \frac{nq}{(nq-1)^3} -2
\frac{nq}{(nq)^3} \right) \ee For $q_0$ sufficiently large, we may
use the bounds: \be \label{Bounds} \frac{n}{ln(n)^C} < \phi(n) < n
\ee and take the leading order in $q$, supposing $q_0$ is large:
\be \sum_n  \left(\frac{nq}{(nq+1)^3}+ \frac{nq}{(nq-1)^3} -2
\frac{nq}{(nq)^3} \right) \approx \frac{ 2 \pi^4}{15  q^4} \ee Now
we may use the bounds (\ref{Bounds}) to evaluate the sum over $q$:
\be \sum_{q>q_0} \phi(q)  \frac{ 2 \pi^4}{15  q^4} < \sum_{q>q_0}
\frac{ 2 \pi^4}{15  q^3} \ee The last sum can be evaluated exactly
in terms of di-$\Gamma$ functions; to leading order in $1/q_0$, we
obtain: \be \Delta_2 < \frac{ 4 \pi^4}{15 q_0^2} \ee Calculating
the lower bound is somewhat trickier; however, we may approximate
it as an integral.  The integral may be evaluated for integer
values of the constant $C$, so we should round $C$ up and then
calculate.  The result is: \be \frac{1}{2 q_0^2 \ln(q_0)^C } \ee
Note that here we do not expect the coefficient to be captured by
the integral, only the qualitative behavior in $q_0$. Hence we
have, for some constant $\alpha$, \be \frac{\alpha}{q_0^2
\ln(q_0)^C } < \Delta_2 < \frac{ 4 \pi^4}{15 q_0^2} \ee

It remains to calculate $\Delta_1$.  To do so, we will first prove the identity:
\be
\sum_{k=1}^n k \phi(k) = \frac{1}{3} \sum_{k=1}^n  k \mu(k) \left [ \lfloor \frac{n}{k}\rfloor ^3
+\frac{3}{2} \lfloor \frac{n}{k}\rfloor ^2 + \frac{1}{2} \lfloor \frac{n}{k}\rfloor \right]
\ee
where $\mu(k)$ is the Mobius function, which takes on values of $+1$ for prime numbers, 0 for perfect squares, and $-1$ otherwise.
This can be shown by induction in $n$; the base case is true since $\phi(1) = \mu(1) =1$.  For the inductive step, we must show:
\ba
(n+1) \phi(n+1) &=& \frac{1}{3} \sum_{k=1}^n  k \mu(k) \left [ \lfloor\frac{n+1}{k}\rfloor ^3 - \lfloor\frac{n}{k}\rfloor ^3 +\frac{3}{2} \lfloor\frac{n+1}{k}\rfloor ^2 - \frac{3}{2} \lfloor\frac{n}{k}\rfloor ^2 \right.\n
&&\left. + \frac{1}{2} \lfloor\frac{n+1}{k}\rfloor- \frac{1}{2} \lfloor\frac{n}{k}\rfloor \right ] + (n+1) \mu(n+1)
\ea
Now, $ \lfloor\frac{n+1}{k}\rfloor^a - \lfloor\frac{n}{k}\rfloor^a $ vanishes except when $k|n+1$, in which case it gives
\be
\left ( \frac{n+1}{k} \right) ^a - \left ( \frac{n+1}{k} -1\right ) ^a
\ee
Substituting this into the series above, we obtain:
\ba
(n+1) \phi(n+1) &=& \frac{1}{3} \sum_{k|n+1, k<n+1}  k \mu(k) \left [ 3 \left (\frac{n+1}{k}\right) ^2-3 \frac{n+1}{k}+1 +3 \frac{n+1}{k} - \frac{3}{2} +\frac{1}{2} \right ] \n
&&+ (n+1) \mu(n+1) \n
&=& (n+1)^2 \sum_{k|n}^n  \frac{ \mu(k)}{k} \n
&=& (n+1)\phi(n+1)
 \ea
where the last equality is a basic identity of Totient functions.
(Proved on Wikipedia).

Now we may use the fact that
\be
\frac{n}{k} -1 < \lfloor\frac{n}{k}\rfloor \leq \frac{n}{k}
\ee
to put bounds on the series.  The upper bound is:
\be
\frac{n^3}{3}  \sum_{k=1}^n  \frac{\mu(k)}{k^2} + \frac{n^2}{2} \sum_{k=1}^n \frac{\mu(k)}{k} + \frac{n}{6} \sum_{k=1}^{n} \mu(k)
\ee
while the lower bound is:
\be
\frac{n^3}{3}  \sum_{k=1}^n  \frac{\mu(k)}{k^2} - \frac{n^2}{2} \sum_{k=1}^n \frac{\mu(k)}{k} + \frac{n}{6} \sum_{k=1}^{n} \mu(k)
\ee
To estimate these contributions, we use the relations:
\ba
\sum_{k=1}^n \frac{\mu(k)}{k^2} &=& \frac{6}{\pi^2} + \mathcal{O} (1/n ) \n
\left| \sum_{k=1}^n \frac{\mu(k)}{k^2}\right| &<& \log(n) +1 \n
\left | \sum_{k=1}^n  \mu(k)\right | &< & n
\ea
Hence at large $n$, we have
\be
\sum_{k=1}^n k \phi(k) =n^3\left (  \frac{6}{\pi^2} + \mathcal{O}(\log(n)/n) \right )
\ee
Plugging this into the expression for $\Delta_1$ above, we have
\ba
\Delta_1 &=& 2 t \sum_{q=1}^{q_0} q \phi(q) \n
\Delta_1&<& \frac{4 t q_0^3}{\pi^2}\left (1 + \frac{\alpha \log(q_0)}{q_0} + \mathcal{O} \frac{1}{q_0}\right)  \n
\Delta_1&>& \frac{4 t q_0^3}{\pi^2}\left (1 - \frac{\alpha \log(q_0)}{q_0} + \mathcal{O} \frac{1}{q_0}\right)
\ea
Plugging in for $t$ in terms of $q_0$, and keeping only the leading term, we obtain:
\be
\Delta_1 \approx \frac{4 \pi^2}{15 q_0 ^2}
\ee

Adding up both contributions, we obtain the following asymptotic form for the volume of the
liquid states in $\mu$:
\be
\frac{\beta}{q_0^2 \ln(q_0)^C} +\frac{4\pi^2}{15 q_0^2}\left (1 - \frac{\alpha \log(q_0)}{q_0}\right)< \Delta_1+\Delta_2 < \frac{4 \pi^4}{15q_0^2}+\frac{4 \pi^2}{15 q_0 ^2} \left (1 + \frac{\alpha \log(q_0)}{q_0} \right)
\ee
We can see that we expect the result to scale approximately as $\frac{1}{q_0^2}$, or as $t^{2/5}$.

\subsection{Creation of double occupancies in CGS states: calculation of
dipolar and quadrupolar interactions} \label{DipoleDef}

Here we discuss the formation of double occupancies in the CGS.  We will
make use of the fact that
 a double occupancy can
always be formed by {\em pinching} about a hole in the CGS -- that is, by moving some number of particles on the
right of that hole leftwards, and some number of particles to the left
of that hole rightwards.

Recall that by even states, we mean states in which either $p$ or $q$ is even; states for which both are odd are odd states.  In an even state, the pinching is
symmetrical; in an odd state it cannot be.  This produces a quadrupolar interaction between
DO in even states, but a dipolar interaction for odd states.

The pinching operation described in Sect. \ref{UStates} can be described by:
 \ba
 R_i \rightarrow R_i -a_i \n
  L_j \rightarrow  L_j - b_j
\ea
where $R_i$, $L_j$ are the positions of the particles relative to
the hole in the initial configuration from which the double occupancy
is formed.  (For an odd state, $R_i$ has one more element
than $L_i$.)

Now let us consider interactions between double occupancies.
The interaction energy between two DO formed a
distance $d$ apart is given by the difference in energies
of the configuration before and after all distances $R_i, L_i$
have been altered, minus the difference in energy when only
a single DO has been formed.  The relevant terms are
the interaction energies between particles that have been pinched
to form the first defect with particles that have been pinched
to form the second:
 \ba
\sum_{i, j}& -[ 1/(d-R_i-L_j +b_i)^3+ 1/(d-R_i-L_j +a_j)^3 ] \n
 & - 1/(d - R_i -L_j +a_i +b_j)^3 - 1/(d-R_i-L_j )^3 ] \n
 & - [1/(d+R_i+L_j +b_i)^3 +1/(d+R_i+L_j +a_j)^3\n
  & - 1/(d + R_i +L_j +a_j +b_i)^3 -  1/(d+R_i+L_j )^3 \n
  &+[1/(d+R_i-R_j+b_j-b_i)^3 - 1/(d+R_i-R_j-b_i)^3 \n
&-1/(d+R_i-R_j +b_j)^3 + 1/(d+R_i-R_j)^3 ]\n
 & +[1/(d+L_i-L_j+a_j-a_i)^3 -1/(d+L_i-L_j-a_i)^3 \n
 & -1/(d+L_i-L_j +a_j)^3 + 1/(d+L_i-L_j)^3 ]
 \ea

For the even case, $R_i =L_i$ and $a_i =b_i$, giving
 \ba
\sum_{i, j} & -[ 1/(d-R_{ij} +a_i)^3 + 1/(d-R_{ij} +a_j)^3 -1/(d -
R_{ij} +a_i+a_j )^3 -  1/(d-R_{ij} )^3 ]\n
  &- [ 1/(d+R_{ij} +a_i)^3 + 1/(d+R_{ij} +a_j)^3 -
  1/(d+R_{ij})^3-1/(d + R_{ij} +a_j +a_i)^3 ] \n
   &+2 [1/(d+R_{ij}-2 L_j+a_j-a_i)^3 - 1/(d+R_{ij}-2 L_j-a_i)^3
\n
 &-1/(d+R_{ij}-2 L_j +a_j)^3 + 1/(d+R_{ij}-2L_j)^3 ]
 \ea
 where $R_{ij} = L_i + L_j$.
 Each term in square brackets is negative definite, by convexity.
A straightforward manipulation of the standard definition of convexity shows
that for $a>b$ and any $x$, we have
 \be
 V(x+a) + V(x+b) > V(x+c) +V(x+(a+b-c)) \ \ \ .
 \ee
 Equally, for $1/r^3$
interactions, the last square bracketed term is necessarily less
than the sum of the first two (since these lie 'closer in', if you
will) and hence the overall interaction is repulsive.

For the odd case, again all terms in square brackets are negative
(by convexity).  However, as there are more $R_i$ than $L_i$,
there are more terms in the last 2 lines than in the first two.
(One more term, to be precise).  Since the last 2 lines are
negative, this results in an interaction that is repulsive at long
distances.

At short distances, however, the interaction is dominated by the
first line, which gives a positive contribution (and hence the
double occupancies repel at sufficiently short distance scales).
Certainly, when $max(R_i +L_j) \approx d$, we expect the potential
to be repulsive and the minimum therefore occurs at some $d > d_0
= max(R_i+L_j)$.  The transition is thus to a density of double
occupancies that sit at least $d_0$ sites apart, and the
transition is to a density of double occupancies that is less than
$1/d_0$.

Further, $d_0$ should increase linearly in $ q$. This is easiest
to see by means of example, but basically creating a DO will form
hole-like solitons, by which I mean you can change distances of
$q$ to $q+1$, but not to $q+2$.  To arrive at such a configuration
involves a re-arrangement over $q$ sites to the right of the DO
(and something like $q/2$ to the left, seemingly).

\begin{figure}[htp]
\begin{center}
 \includegraphics[totalheight=10cm]{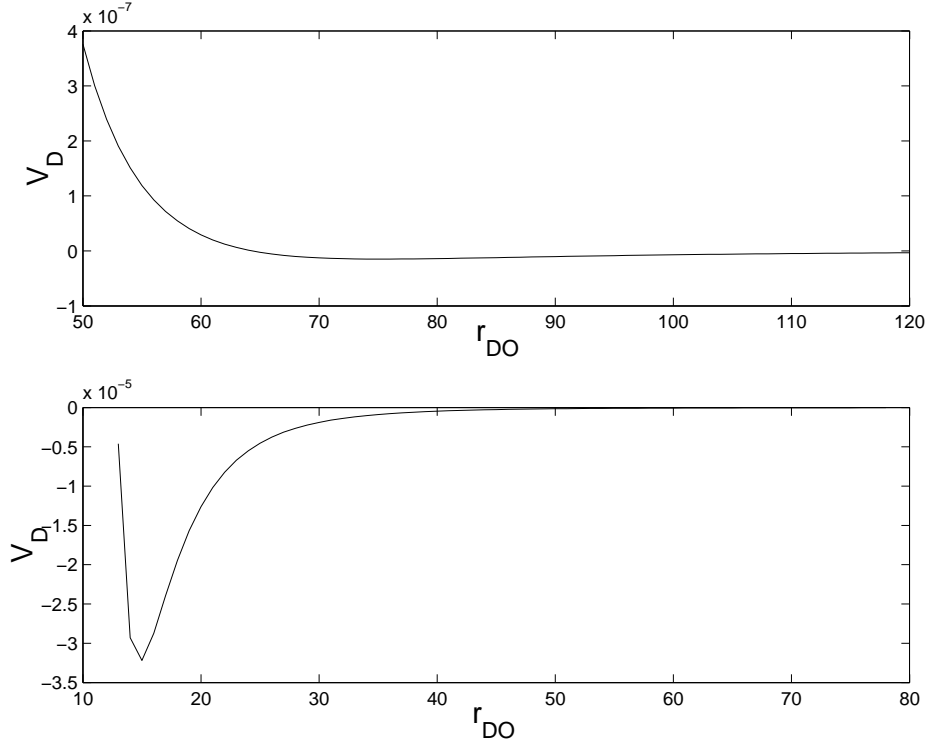}
 \end{center}
\caption[Dipolar interaction of double occupancies in certain odd-denominator filling fractions]{\label{PinchFig}  Two-DO interaction potential $V_{D}$ for the $1/3$ (lower plot) and $1/5$ (upper plot) filled
states.  Here $r_{DO}$ is the distance between the two double occupancies in units of the lattice constant.  The minima of these potentials are at $r_{DO} \approx 75 $ and $15 $, respectively, for the $1/3$ and $1/5$ -filled states.}
\end{figure}

As an example, we plot the $2$ DO interaction potential as a function
of separation between the doubly occupied site of each DO for the $1/3$ and $1/5$
filled states in Fig. \ref{PinchFig}.  The minima sit at $15$ and $75$ lattice spacings,
respectively.

Hence as $U_0$ is lowered, we expect even states to undergo a
second-order phase transition, as the interaction between DO is
quadrupolar and hence repulsive.  For odd states the interaction is
attractive, suggesting a first-order phase transition in which
a finite density of defects forms.  However, this density of
defects vanishes at least as $1/q$, and quite possibly more
rapidly than this.

\subsection{Lattice-scale arguments for adding charge as double occupancies}  \label{DOAdd_App}

Here we present a more complete argument that charges added
as solitons have a stronger mutual repulsion energy than
charges added as double occupancies.  At long length scales we
may employ a simple intuition based on electrostatics.  That is, 
when adding charges to the system we must compare both their self-energy 
(in this case, the energy of adding a single charge) and their interaction.
The interaction energy between $n$ charges $q$, spread over some finite interval 
of length $L$ in the system, is $E(n)~q \left( \frac{n}{L} \right )^3$.  This is 
always less than the interaction energy between 
$m n$ charges of charge $q/m$ for any integer $m>1$,  which is $E(m, n) ~ \frac{q}{m} \left( \frac{Lmn}{L} \right)^3 = m^2 E(n)$.
We work in the r{\'e}gime where the self-energy favors doubly occupied states over 
solitons -- hence the charge will never fractionalize.

Here we outline a lattice argument to show that this intuition 
gives the correct result even at short distances.  We treat only
the case $\nu \geq 1/2$, as the form of the doubly occupied
states is simpler to describe; however, we expect similar
arguments to hold for arbitrary filling fractions.  

Consider adding a finite density of charges to a state 
with existing double-occupancies.  We break the
energy cost of introducing the new particle into two pieces:
\ba
E &=& E_l + E_I
\ea
where $E_l$ is the self-energy, and $E_I$ is the interaction energy.  The
interaction energy contains the repulsion between the extra charge and any existing
double-occupancies.  If the extra charge breaks into solitons, $E_I$ also 
contains repulsion terms between these solitons.  Since we add 
a finite density of charge, we can consider that each charge is effectively 
confined to a finite region of length $r$ in the system.

Consider first the $1/2$-filled state.  If the charge fractionalizes, two solitons
will be formed.  We place these solitons at $r_1$ and $r$,
respectively, from the double occupancy.  Here $r_1$ and $r$
measure the distance from the DO to the outer-most edge of each
soliton.  One soliton sits as far from the DO as possible, as the
repulsion between the DO and each soliton is greater than the
soliton-soliton repulsion, as a DO has integral charge while the
charge of a soliton is fractionalized.  We may think of forming
this arrangement by excising a hole at $r_1$, pushing the rest of
the pattern inwards, and placing an extra particle at $r$. For $r
= 10$, this looks like: \ba 2 0 1 0 1 {\bf 0} 1 0 1 0 1 \n 2 0 1 0
1 1 0 1 0 1 {\bf 0} \n 2 0 1 0 1 1 0 1 0 1 {\bf 1} \ea Note that
$r_1$ is necessarily odd, and $r$ here must be even. Then clearly
we have:
 \ba E_I (sol)& =& \mbox{min}_{r_1}
\sum_{n=0}^{(r-r_1-1)/2} \left [ \frac{1}{(r_1 + 2 n)^3}  -
\frac{1}{(r_1 + 2n +1)^3} \right ] + \frac{1}{r^3} \n
 & &+ 2
\sum_{n=0}^{\infty}\left[  \frac{1}{ (r-r_1 + 2n)^3 }
-\frac{1}{(r-r+1 +2n +1)^3} \right. \n
 && \left.
 + \frac{1}{(r-r_1 + 2n +1)^3} - \frac{1}{ (r-r_1 +2n +2)^3} \right ]
\ea
The first line here is the interaction between the soliton and the DO; the second
line gives the soliton-soliton repulsion.  (One pair of terms for each particle
in the soliton).  The sums are all positive, since the $1/r^3$ potential is
monotonically decreasing.  Hence, the interaction energy from adding
a particle as solitons is strictly greater than that for adding it
as a double occupancy:
\be
E_I (sol) > \frac{1}{r^3 } = E_I (DO) \ \ .
\ee

Since $r$ above is arbitrary, this argument also applies to the
case of adding a particle to a system with multiple double
occupancies in the $1/2$-filled state.  Hence in any finite
system, we can induct on the number of particles to show that
every filling $p/q>1/2$ consists only of double occupancies in the
$1/2$ filled state, and never of soliton-like insertions $... 0 1
1 0 ...$. In an infinite system, an infinite number of particles
must be added at once in order to change the filling fraction; in
this case the particles want to spread out homogeneously and hence
we again confine them to within some distance $r$ of existing
double occupancies.  (Similar arguments show that the repulsion
between four solitons in a finite region is greater than the
repulsion between two double-occupancies, as the latter can spread
farther apart).

Finally, we may generalize this argument to arbitrary fillings
$p/q > 1/2$.  Again we form solitons by pushing a hole from some
radius $r_1$ to the farthest possible radius $r$ from the DO, and
inserting an extra particle at $r$.  This is illustrated below for
the $2/3$ filled state:
 \ba
 2 0 1 1 0 1 1 {\bf 0} 1 1 0 1 1 0 1 1  \n
 2 0 1 1 0 1 1 1 1 0 1 1 0 1 1 {\bf 0} \n
 2 0 1 1 0 1 1 1 0 1 1 1 0 1 1 {\bf 1}
 \ea
 We may hop some (but not all) of the particles back to their
original positions to form solitons, as shown in the third line.
The energy of such a state clearly obeys
 \be
  E_I(sol) > \frac{1}{r^3} = E_I(DO)
  \ee
and hence again, all additional particles will enter as double
occupancies.  This shows that the doubly occupied state is 
locally stable at all length scales.

\bibliography{thesis}

\end{document}